\documentclass[iop]{emulateapj}

\usepackage{natbib,aas_macros,amsmath}
\usepackage{multirow,color}

\newcommand{\rqq}{\textquotedblright}
\newcommand{\lqq}{\textquotedblleft}

\newcommand{\carcsec}{$\!\!\arcsec$}
\newcommand{\m}[1]{\mathrm{#1}}
\newcommand{\kms}{\m{km\ s^{-1}}}
\newcommand{\ob}{{\sc Oii}B\ 10}
\newcommand{\Msun}{\ensuremath{M_{\odot}}}

\begin{document}
\shortauthors{Harikane et al.}
\slugcomment{Accepted for publication in the Astrophysical Journal}

\email{hari@icrr.u-tokyo.ac.jp}
\author{%
Yuichi Harikane\altaffilmark{1,2}, Masami Ouchi\altaffilmark{1,3}, Suraphong Yuma\altaffilmark{1}, Michael Rauch\altaffilmark{4}, Kimihiko Nakajima\altaffilmark{5}, Yoshiaki Ono\altaffilmark{1}
}

\altaffiltext{1}{%
Institute for Cosmic Ray Research, The University of Tokyo, 5-1-5 Kashiwanoha, Kashiwa, Chiba 277-8582, Japan
}
\altaffiltext{2}{%
Department of Physics, Graduate School of Science, The University of Tokyo, 7-3-1 Hongo, Bunkyo, Tokyo, 113-0033, Japan
}
\altaffiltext{3}{%
Kavli Institute for the Physics and Mathematics of the Universe (Kavli IPMU), WPI, The University of Tokyo, 5-1-5, Kashiwanoha, Kashiwa, Chiba 277-8583, Japan
}
\altaffiltext{4}{%
Carnegie Observatories, 813 Santa Barbara Street, Pasadena, CA 91101, USA
}
\altaffiltext{5}{%
National Astronomical Observatory of Japan, 2-21-1 Osawa, Mitaka, Tokyo 181-8588, Japan
}

\shorttitle{Spectroscopy for an [OII] Blob at $\lowercase{z}=1.18$}

\title{%
MOSFIRE and LDSS3 Spectroscopy for an [OII] Blob at $\lowercase{z}=1.18$:\\
Gas Outflow and Energy Source
}

\begin{abstract}
We report our Keck/MOSFIRE and Magellan/LDSS3 spectroscopy for an {\sc [Oii]} Blob, {\ob},
that is a high-$z$ galaxy with spatially extended {\sc [Oii]}$\lambda\lambda3726,3729$ emission over 30 kpc
recently identified by {a} Subaru large-area narrowband survey.
The systemic redshift of {\ob} is $z=1.18$ securely determined with {\sc [Oiii]}$\lambda\lambda4959,5007$ and
H$\beta$ emission lines. 
We identify Fe{\sc ii}$\lambda$2587 and Mg{\sc ii}$\lambda\lambda$2796,2804 absorption lines blueshifted from the systemic redshift by $80\pm50$ and $260\pm40$ km s$^{-1}$, respectively, which indicate gas outflow from {\ob} with the velocity of $\sim 80-260$ km s$^{-1}$.
This outflow velocity is comparable with the escape velocity, $250\pm140$  km s$^{-1}$,
estimated under the assumption of a singular isothermal halo potential profile.
Some fraction of the outflowing gas
could escape from the halo of {\ob}, suppressing {\ob}'s star-formation activity.
We estimate a mass loading factor, $\eta$, that is a ratio of mass outflow rate to star-formation rate, and obtain $\eta>0.8\pm 0.1$ which is relatively high compared with low-$z$ starbursts including U/LIRGs and AGNs.
The major energy source of the outflow is unclear with the available data.
Although no signature of AGN is found in the X-ray data, 
{\ob} falls in the AGN/star-forming composite region in the line diagnostic diagrams.
It is possible that the outflow is powered by star formation and a type-2 AGN
with narrow FWHM emission line widths of $70-130$  km s$^{-1}$.
This is the first detailed spectroscopic study {of} oxygen-line blobs{, which includes the analyses of the escape velocity, the mass loading factor, and the presence of an AGN}, and a {significant} step
to understanding the nature of oxygen-line blobs and the relation with gas outflow and star-formation
quenching at high redshift.
\end{abstract}


\section{Introduction}

Galactic outflows are thought 
to play a significant role in galaxy formation and evolution. 
Theoretical studies to reproduce 
the observed luminosity function of galaxies 
have claimed the need for physical mechanisms 
that are able to modulate the efficiency of star formation 
\citep[e.g.,][see also the review of \citealt{2006RPPh...69.3101B}]{2000MNRAS.319..168C,2003MNRAS.339..312S,2009MNRAS.396.2332K}.
One of the most popular proposed mechanisms for regulating the star forming activity 
is a galactic outflow driven by active galactic nuclei (AGNs) and/or star formation (followed by supernova explosions, stellar winds, and radiation pressure). 
The AGN and star formation feedback processes are thought to be 
necessary for explaining the shape of the galaxy luminosity function 
at the bright and faint ends, respectively 
\citep[e.g.,][]{2005ApJ...618..569M,2008MNRAS.391..481S}. 
Outflows would also be a primary mechanism for two important processes.
One is the chemical enrichment of the circumgalactic medium (CGM) and the intergalactic medium 
\citep[IGM; e.g.,][]{2005ApJ...621..227M, 2005ApJS..160...87R, 2005ApJS..160..115R, 2009ApJ...692..187W, 2011ApJ...743...46C}.
The other is forming the mass-metallicity relation of galaxies 
\citep[e.g.,][]{2004ApJ...613..898T,2006ApJ...644..813E}. 
Furthermore, 
outflows may play an important role in cosmic reionization 
by opening up low-density paths 
for ionizing photons to escape from star-forming galaxies \citep{2000ApJ...531..846D,2001ApJ...558...56H,2002MNRAS.337.1299C,2003ApJ...599...50F,2008ApJ...672..765G,2009MNRAS.398..715Y,2010ApJ...710.1239R}. 

Various observational studies have found 
evidence for the presence of galactic-scale outflows  
in local starburst galaxies, 
 including 
 dwarf galaxies \citep[e.g.,][]{1995A&A...301...18L,1997ApJ...482..114H,2001ApJ...554.1021H,2004ApJ...610..201S} and ultra-luminous infrared galaxies \citep[ULIRGs; e.g.,][]{2000ApJS..129..493H,2002ApJ...570..588R,2005ApJ...621..227M,2009ApJ...692..187W}. 
These outflows are mostly traced via the blueshifts of interstellar 
absorption lines with respect to the galaxy systemic redshifts.
%
At high redshifts, 
outflows in star-forming galaxies have been identified by 
the detection of blueshifted interstellar absorption lines \cite[e.g.,][]{2002ApJ...569..742P,2003ApJ...588...65S,2007ApJ...663L..77T,2010ApJ...717..289S,2011ApJ...730....5H,2011ApJ...743...46C,2012ApJ...759...26E,2012ApJ...760..127M,2012ApJ...758..135K,2012ApJ...751...51J,2013ApJ...765...70H,2014arXiv1402.1168S}.
Observational studies as well as theoretical studies 
have suggested that there exists 
a critical star-formation rate (SFR) surface density of 
$\sim 0.1 M_\odot$ yr$^{-1}$ kpc$^{-2}$ 
to launch galactic-scale outflows \citep{2002ApJ...577..691H,2011ApJ...735...66M,2012ApJ...758..135K}.
Outflows appear to be ubiquitous 
in both nearby and high-$z$ galaxies 
with such intense star formation activities.

Although 
outflow signatures have been unambiguously detected in many systems, 
absorption line detections offer limited information on the spatial distribution of the outflowing gas. 
The spatial distribution of outflowing gas can be measured, 
if galactic outflows can be observed in the spatially extended emission. 
In fact, in the last fifteen years,  
a growing number of observational studies have reported 
their discoveries of extended {\sc Hi} Ly$\alpha$ nebulae at high redshifts 
\citep[e.g.,][]{1999MNRAS.305..849F,2000ApJ...532..170S,2001ApJ...554.1001F,2004ApJ...602..545P,2004AJ....127.1313O,2004AJ....128..569M,2006A&A...452L..23N,2006ApJ...648...54S,2007MNRAS.378L..49S,2009ApJ...693.1579Y,2009ApJ...696.1164O,2011MNRAS.410L..13M,2011ApJ...736..160S,2012ApJ...752...86P,2014arXiv1403.0732M}.
Several mechanisms have been proposed 
to explain the extended Ly$\alpha$ nebulae 
including strong outflows driven by AGNs and/or star formation \citep[e.g.,][]{2000ApJ...532L..13T,2003ApJ...591L...9O,2004AJ....128..569M,2004ApJ...613L..97M,2004MNRAS.351...63B,2005Natur.436..227W,2006ApJ...637L..89C,2009ApJ...692.1561W,2010MNRAS.402.2245W}.
However, 
a photoionization by AGNs and/or an intense star formation
is also a plausible origin of the extended Ly$\alpha$ \citep[e.g.,][]{2005ApJ...629..654D,2009ApJ...700....1G,2013ApJ...771...89O}.
Since Ly$\alpha$ is a resonance line,
the extensive distributions of Ly$\alpha$ emission may be 
produced by the resonance scattering of Ly$\alpha$ photons in the CGM \citep{2007ApJ...657L..69L,2011Natur.476..304H,2011ApJ...739...62Z,2012MNRAS.424.1672D,2012A&A...546A.111V,2012MNRAS.424.2193J}. 
Conceivably, 
an inflowing gas stream into the dark matter halo 
could also contribute to the extended Ly$\alpha$ emission \citep[e.g.,][]{2000ApJ...537L...5H,2001ApJ...562..605F,2006MNRAS.368....2D,2006ApJ...649...37D,2009MNRAS.400.1109D,2010MNRAS.407..613G,2014arXiv1403.0732M}.


{Unlike the resonant Ly$\alpha$ line, the spatially extended metal emission lines are more straightforwardly related to galactic outflows.}
{Inspired by the method of \citet{2004AJ....128..569M} who have conducted the systematic search for galaxies with extended Ly$\alpha$ emission,} \citet[hereafter Y13]{2013ApJ...779...53Y} have used wide and deep narrowband images of the Subaru/Suprime-Cam to systematically search for galaxies with extended emission of non-resonant metal lines.
Y13 have identified $12$ star-forming galaxies at $z\sim1.2$ 
whose [{\sc Oii}]$\lambda\lambda3726,3729$ emission extend over $30$ kpc. 
These spatially extended [{\sc Oii}]-emitting galaxies, 
which they call {\lqq}[{\sc Oii}] blobs ({\sc Oii}Bs){\rqq}, 
offer a unique laboratory for 
investigating the physical properties of galactic-scale outflows, 
and 
give clues to understanding the role of the feedback 
in galaxy formation and evolution. 


In Y13, 
three out of the $12$ {\sc Oii}Bs ({\sc Oii}B\ 1, 4, and 8) 
have been spectroscopically observed in the optical wavelength.  
Y13 have reported that 
blueshifted interstellar absorption lines are detected 
in the rest-frame UV spectra of {\sc Oii}B\ 1 and 4\footnote{Y13 
detect no UV continuum of {\sc Oii}B\ 8 due to the poor signal-to-noise ratio of their spectrum.}, 
which is indicative of outflows. 
Furthermore, 
Y13 have found that {\sc Oii}B\ 1 is the largest and brightest in their sample, and classified {\sc Oii}B\ 1 as a radio-quiet obscured type-2 AGN  
based on the detection of [Ne {\sc v}] emission, 
a broad [{\sc Oii}] line width, a high [{\sc Oii}] equivalent width (EW), 
and the ratio of mid-infrared to radio fluxes.  
This indicates that 
the outflow and the spatially extended [{\sc Oii}] emission of {\sc Oii}B\ 1 
would be powered by the AGN activity. 
{However, the fraction of {\sc Oii}Bs having AGN signatures is unknown.
Thus}, it is not clear whether 
the extended [{\sc Oii}] emission of the {\sc Oii}B population 
and their possible outflows 
are mainly driven by their AGN activities 
or powered by other energy sources such as star formation.

In this study, we focus on {\ob} ($\m{R.A.=2^{h}17^{m}32{\fs}53}$, 
$\m{dec.=-4{\arcdeg}57{\arcmin}46{\farcs}40}$) originally identified as an {\sc Oii}B by Y13.  
{We choose \ob\ for our target for two reasons.
The first reason is that \ob\ is expected to have an outflow whose major heating source would be star formation.
In the available data of \ob, no signature of AGN is found.
Moreover, \ob\ is one of the {\sc Oii}Bs with the highest specific star formation rate (SSFR).
The second reason is that \ob\ is observed as a mask filler for the other program.
We describe this observation in Section \ref{obda_mos}.}
The aim of this study is 
to characterize the properties of \ob\ 
based on our deep optical and near-infrared spectroscopic observations 
as well as archival multi-wavelength imaging data.  
These observational data enable us to 
examine whether an outflow is occurring in {\ob} 
and to investigate the physical origin of the extended [{\sc Oii}] nebula of {\ob}.
This is the first detailed spectroscopic study {of} oxygen-line blobs {which includes the analyses of the escape velocity, the mass loading factor, and the presence of an AGN, and is a significant }step
toward understanding the nature of oxygen-line blobs and the relationship between gas outflow and star formation
quenching at high redshift.

This paper is organized as follows. We present the target selection and photometric data in Section \ref{pd}. We describe Keck/MOSFIRE and Magellan/Low-Dispersion Survey Spectrograph (LDSS3)  observations and data analyses in Section \ref{obdr}. The results are shown in Section \ref{a}. We discuss the nature of \ob\ in Section \ref{r}. Section \ref{s} summarizes our findings. Throughout this paper, magnitudes are in the AB system, and we assume a standard $\m{\Lambda}$ CDM cosmology with parameters of $(\Omega_{\rm m},\Omega_\Lambda,H_0)=(0.3,0.7,70\ \m{km\ s^{-1}\ Mpc^{-1}})$. 
In this cosmology, 
an angular dimension of 1.0 arcsec corresponds to 
a physical length of 8.32 kpc at $z=1.18$.

\section{Target Selection and Multi-wavelength Imaging Data} \label{pd}

\subsection{Target Selection for the Spectroscopic Observations}\label{tsso}

\begin{figure}
\begin{center}
\includegraphics[width=90mm,bb=30 50 420 420]{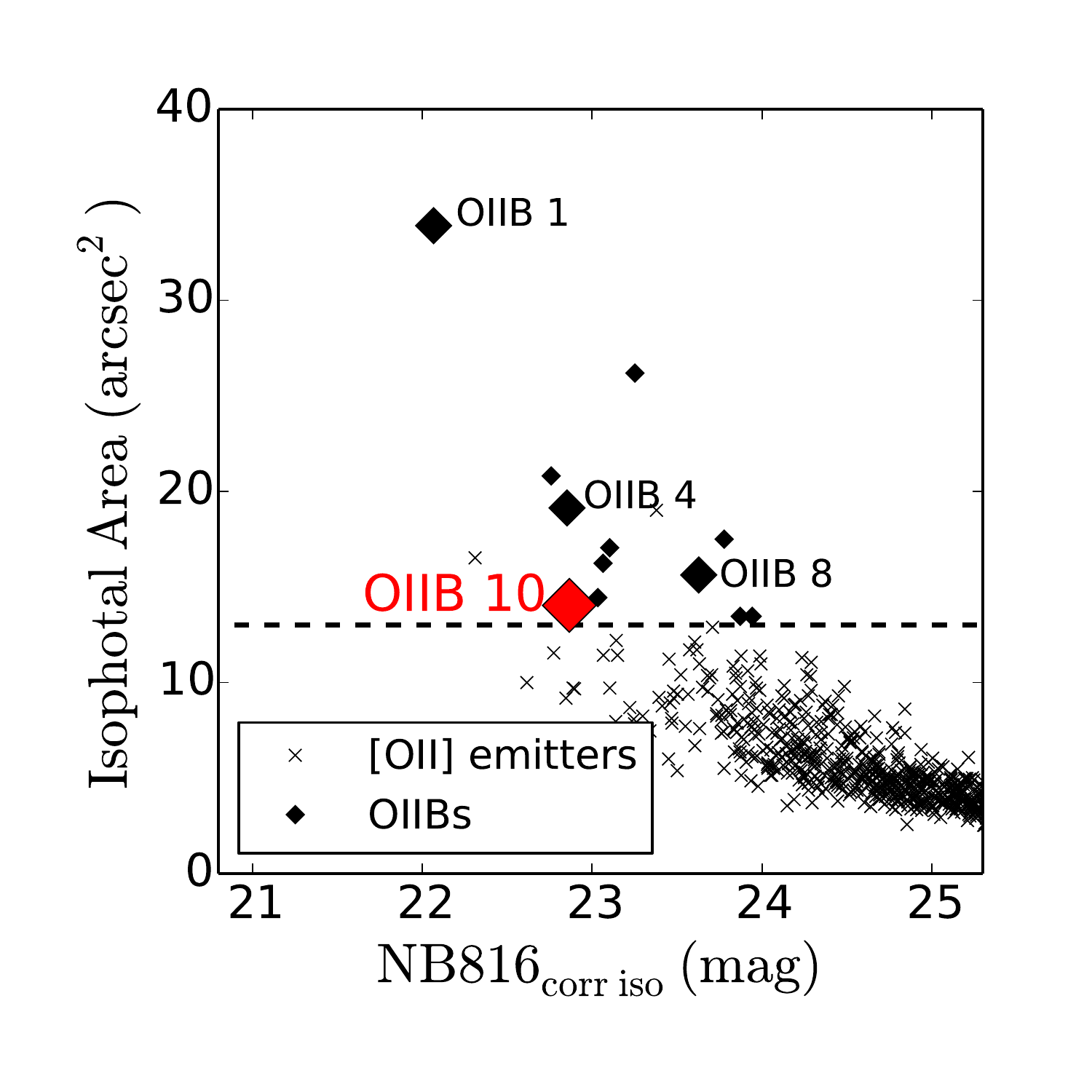}
\end{center}
\caption{Distribution of isophotal area and magnitudes in the ${\it{NB816_\m{corr}}}$ image. The isophotal area criterion ($>13\ \m{arcsec^2}$) is shown with the horizontal dashed line. The red diamond shows our target, \ob. The black diamonds represent the rest of 11 {\sc Oii}Bs, and black crosses denote {\sc [Oii]} emitters found by \citet{2013MNRAS.433..796D}. {The two black crosses above the dashed line are {\sc [Oii]} emitters that are not {\sc Oii}Bs, but objects with a large isophotal area made by source blending.} \label{fig_mag_isoarea}}
\end{figure}

Y13 have identified twelve {\sc Oii}Bs using a catalog of {\sc [Oii]} emitters at $z\sim1.2$ \citep{2013MNRAS.433..796D}. {These {\sc [Oii]} emitters show strong {\sc [Oii]} emission lines falling into} the narrowband filter $NB816$ with the central wavelength of $8150\ \m{\AA}$ and an FWHM of $120\ \m{\AA}$. To select {\sc Oii}Bs {from the normal {\sc [Oii]} emitters}, Y13 have made the emission-line $NB816_{\m{corr}}$ image by subtracting a continuum-emission image from the $NB816$ image. 
The continuum-emission image, dubbed "$Rz$ image", is constructed from the $R$ and $z$ band images.
Y13 define the isophotal area as pixels with values above the 2$\sigma$
sky fluctuation ($28\ \m{mag\ arcsec^{-2}}$ or $1.2\times10^{-18}\ \m{erg\ s^{-1}\ cm^{-2}\ arcsec^{-2}}$) of the $NB816_{\m{corr}}$ image.
Following the procedures in \citet{2004AJ....128..569M}, Y13 classify an object as a candidate of {\sc Oii}B if its isophotal area is above $13\ \m{arcsec^2}$ (corresponding to a spatial extent of $30\ \m{kpc}$ at $z\sim1.2$) in the $NB816_{\m{corr}}$ image.
After visual inspection, twelve candidates are identified as {\sc Oii}Bs.
More details are described in Y13.
From twelve {\sc Oii}Bs, we select \ob\ for our target of the spectroscopic observations, because \ob\ is expected to have an outflow driven by its star-formation activity.
\ob\ has no signature of AGN in the available data, and the SSFR of \ob\ is relatively higher than those of other {\sc Oii}Bs. 
The isophotal magnitude and area of \ob\ in the $NB816_{\m{corr}}$ image are $22.89\ \m{mag}$ and $14\ \m{arcsec^2}$, respectively, as presented with the red diamond in Figure \ref{fig_mag_isoarea}.

\subsection{Multi-wavelength Imaging Data and Stellar Population of {\ob}\label{mw}}

\begin{figure*}
 \begin{minipage}{0.19\hsize}
  \begin{center}
   \includegraphics[width=30mm]{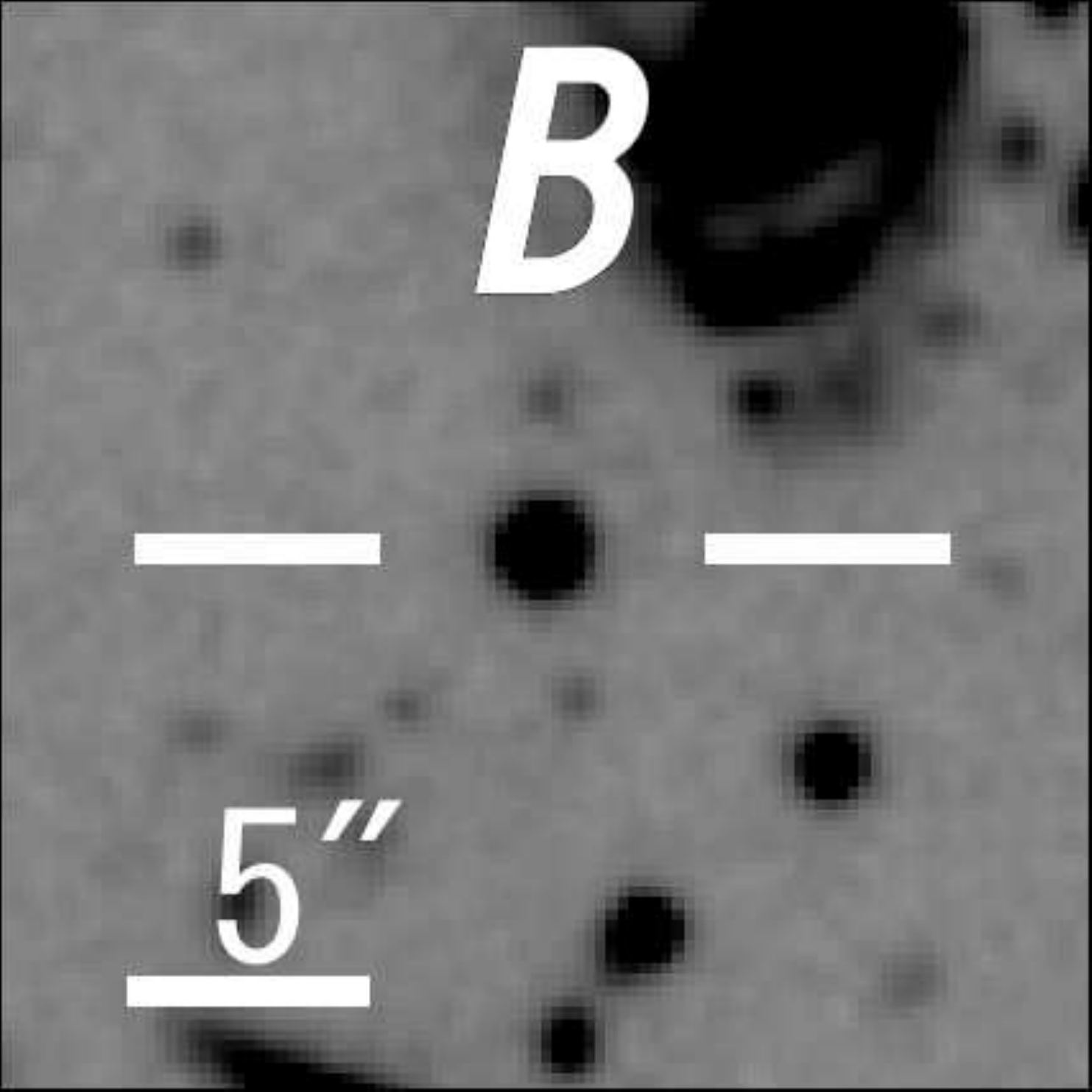}
  \end{center}
  \label{fig:one}
 \end{minipage}
 \begin{minipage}{0.19\hsize}
 \begin{center}
  \includegraphics[width=30mm]{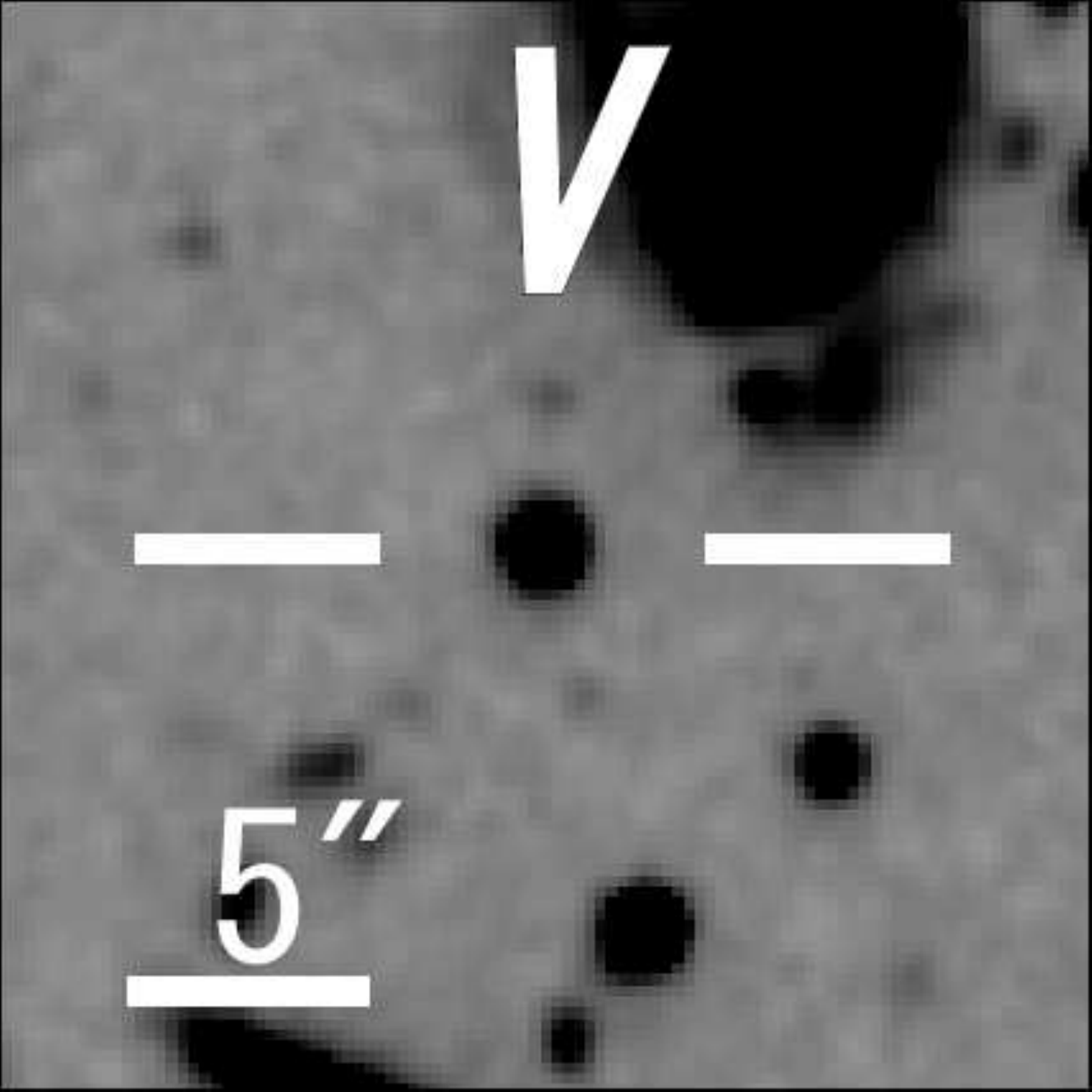}
 \end{center}
  \label{fig:two}
 \end{minipage}
 \begin{minipage}{0.19\hsize}
 \begin{center}
  \includegraphics[width=30mm]{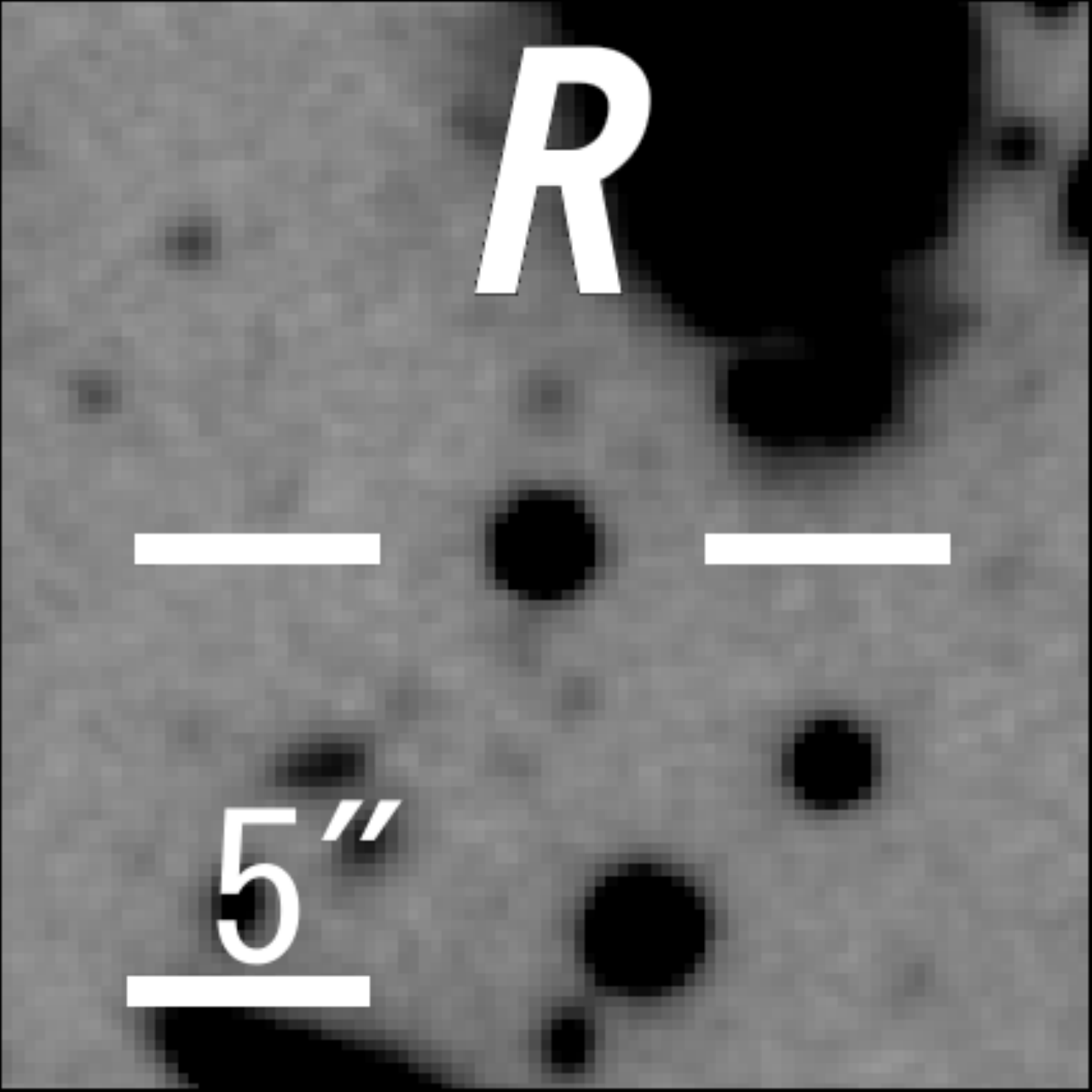}
 \end{center}
  \label{fig:three}
 \end{minipage}
  \begin{minipage}{0.19\hsize}
 \begin{center}
  \includegraphics[width=30mm]{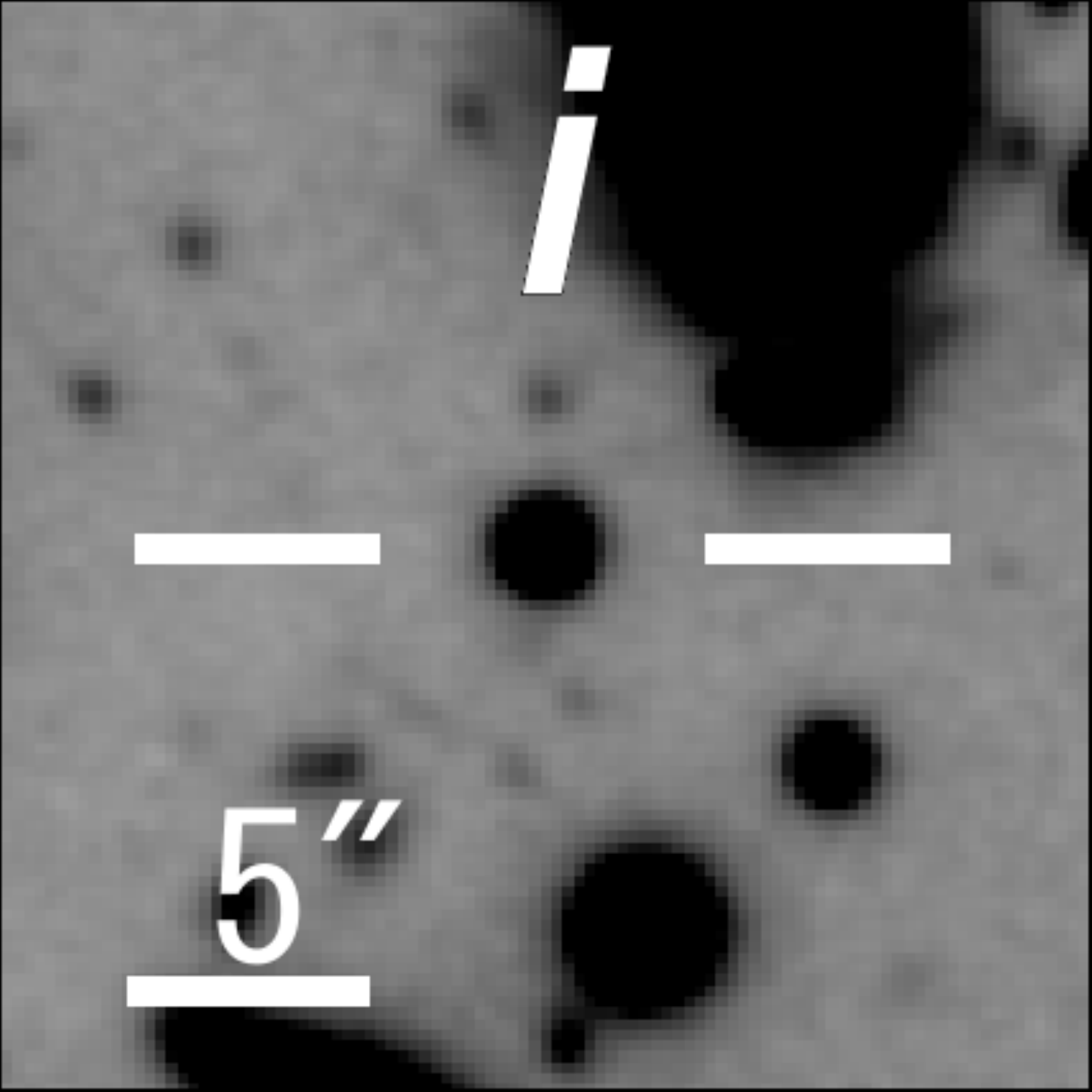}
 \end{center}
  \label{fig:three}
 \end{minipage}
  \begin{minipage}{0.19\hsize}
 \begin{center}
  \includegraphics[width=30mm]{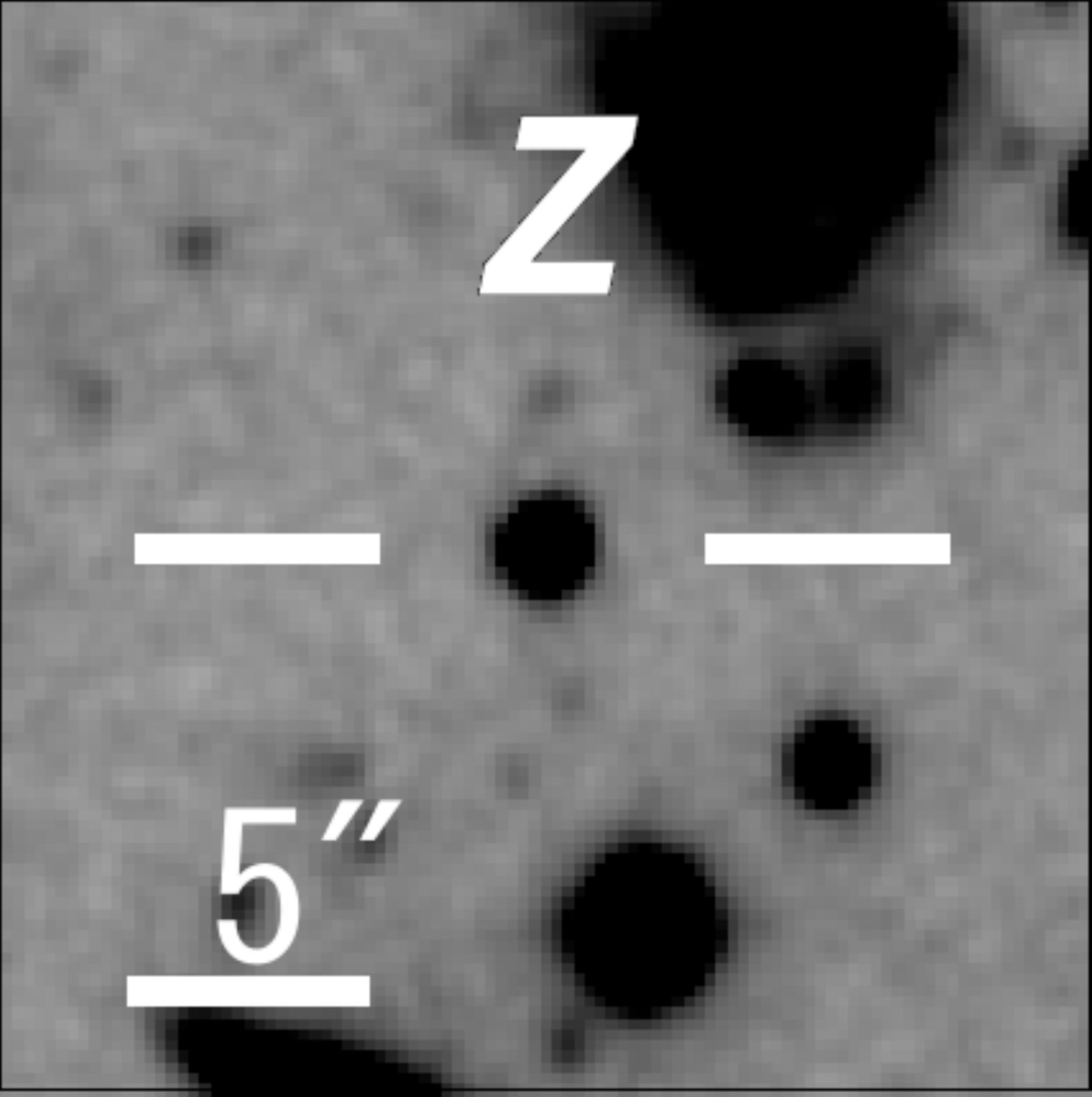}
 \end{center}
  \label{fig:three}
 \end{minipage}
 
  \begin{minipage}{0.19\hsize}
  \begin{center}
  \includegraphics[width=30mm]{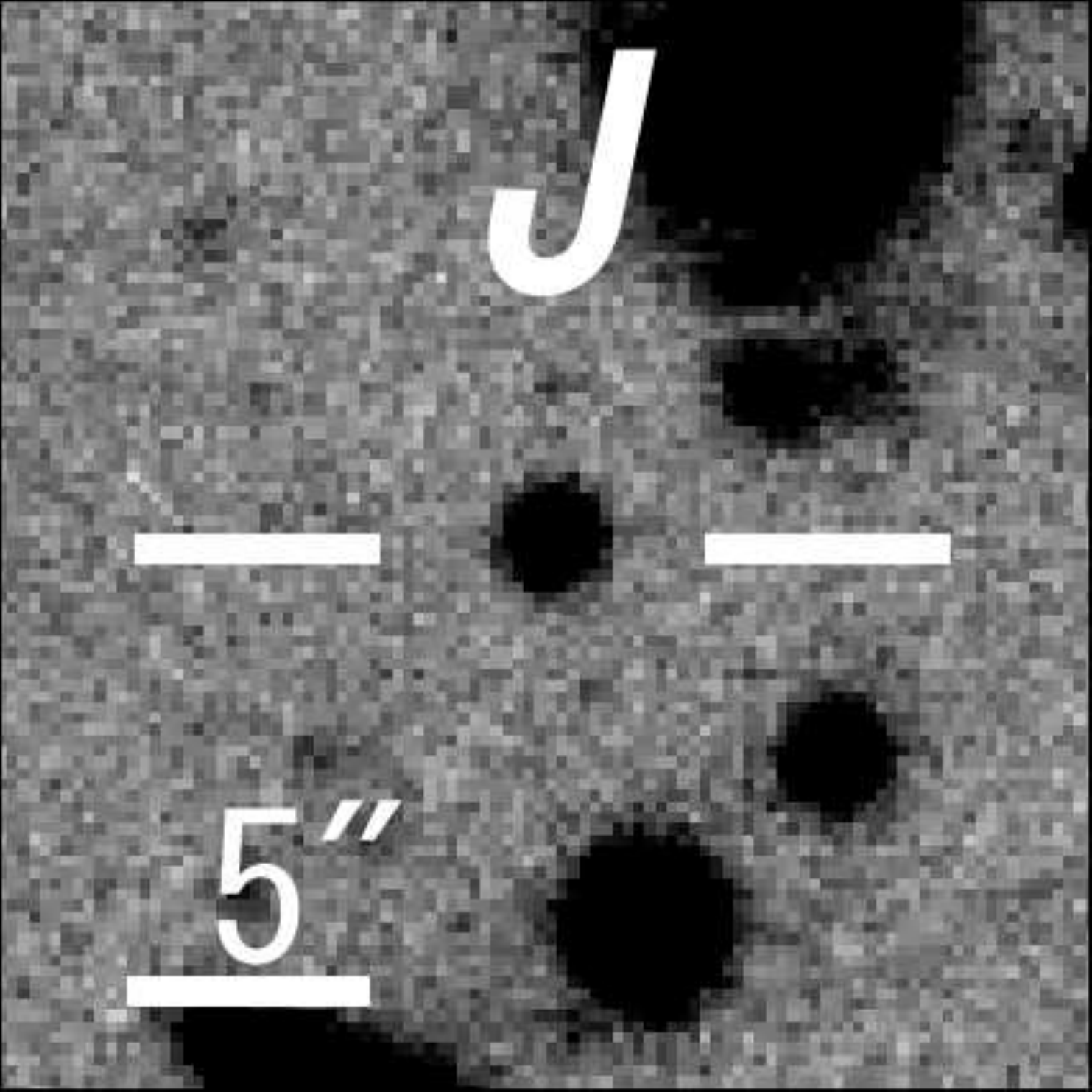}
  \end{center}
  \label{fig:one}
 \end{minipage}
 \begin{minipage}{0.19\hsize}
 \begin{center}
 \includegraphics[width=30mm]{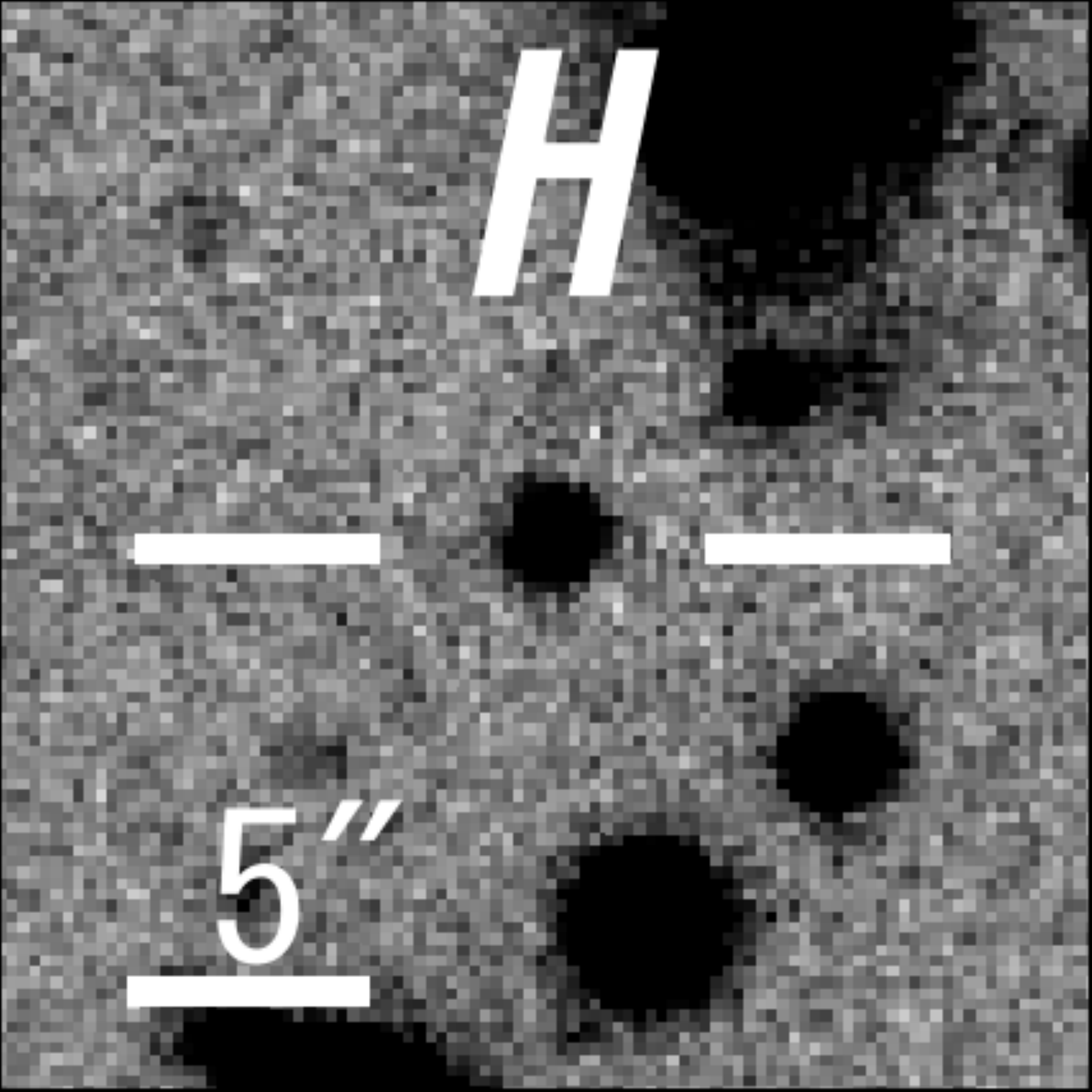}
 \end{center}
  \label{fig:two}
 \end{minipage}
 \begin{minipage}{0.19\hsize}
 \begin{center}
 \includegraphics[width=30mm]{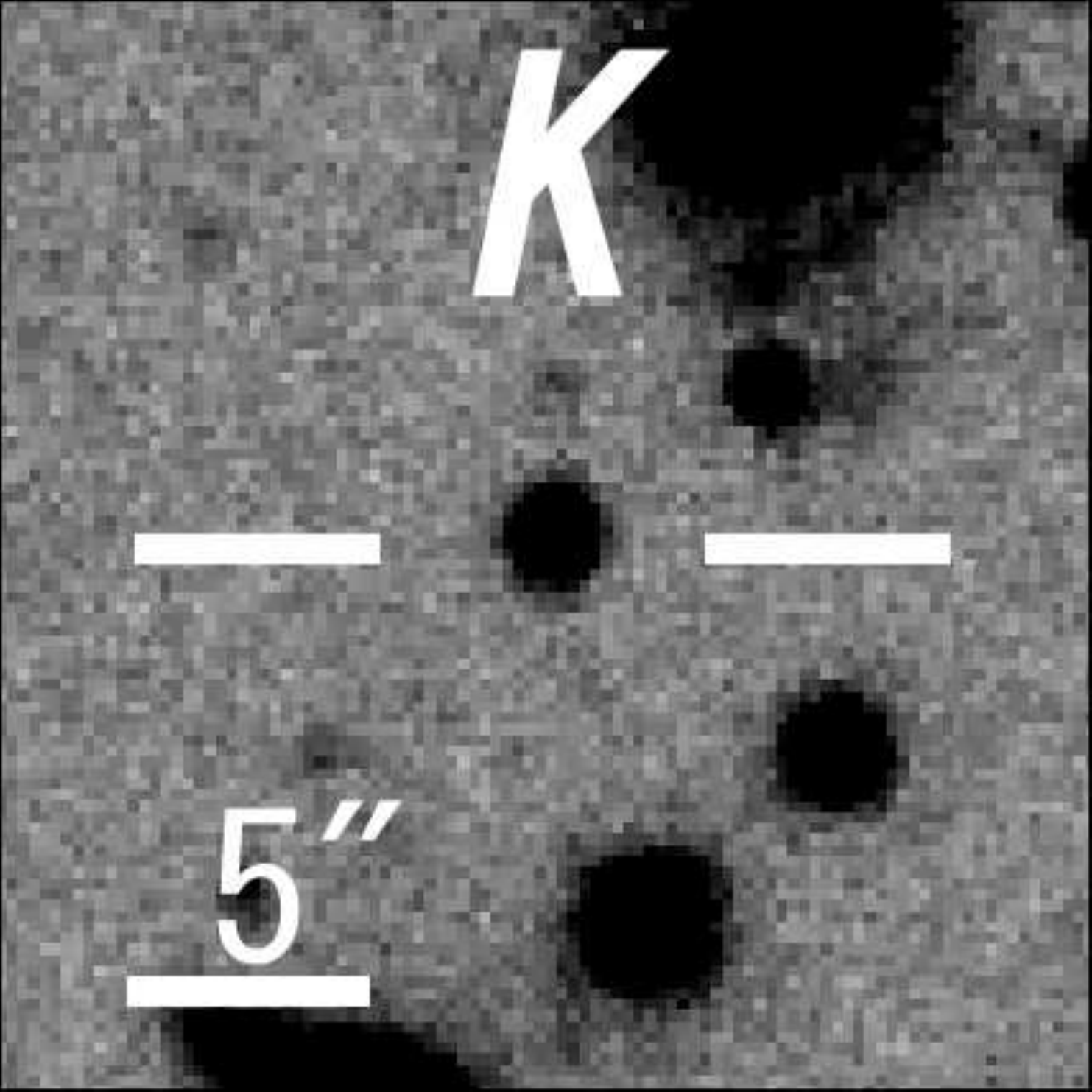}
 \end{center}
  \label{fig:three}
 \end{minipage}
  \begin{minipage}{0.19\hsize}
 \begin{center}
  \includegraphics[width=30mm]{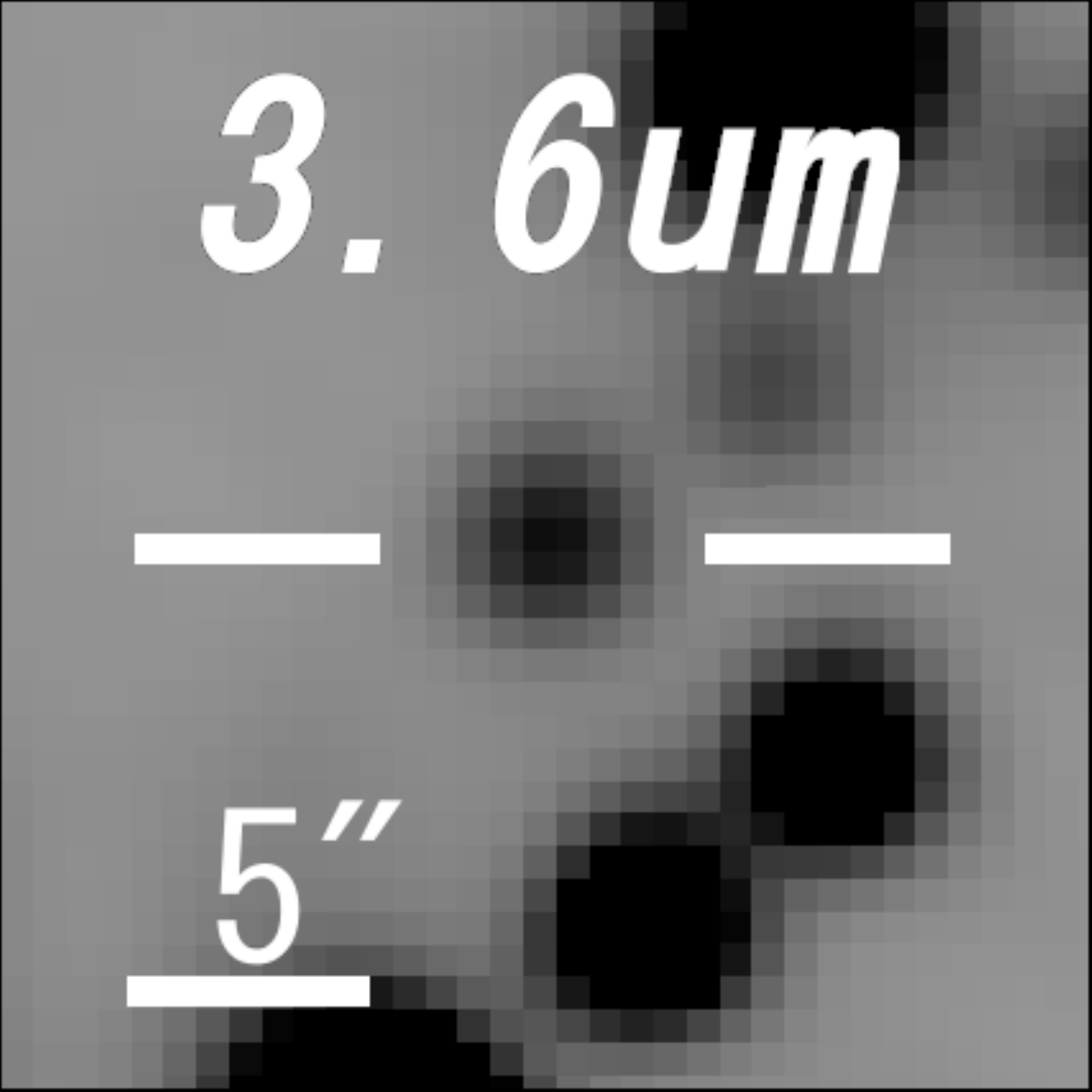}
 \end{center}
  \label{fig:three}
 \end{minipage}
  \begin{minipage}{0.19\hsize}
 \begin{center}
  \includegraphics[width=30mm]{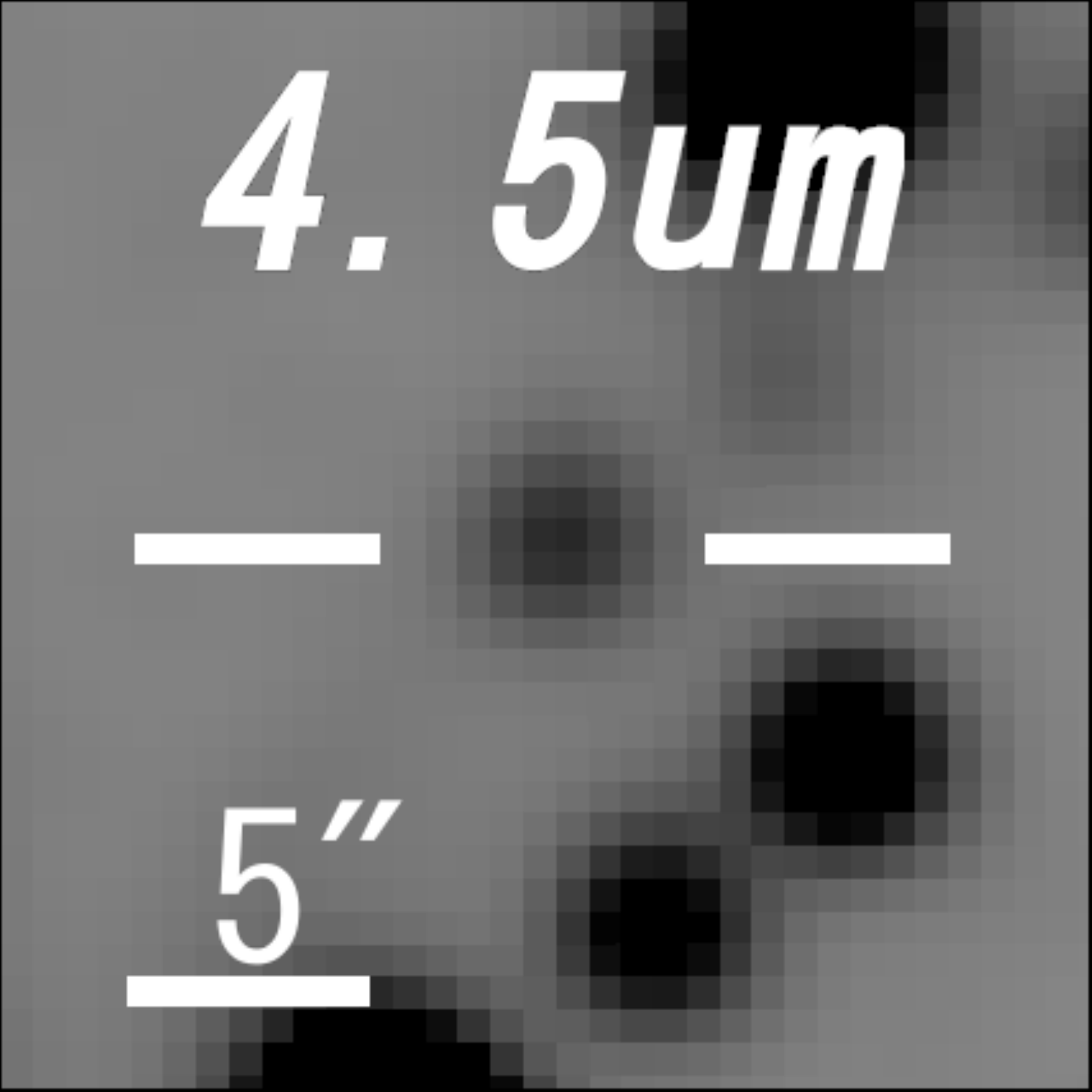}
 \end{center}
  \label{fig:three}
 \end{minipage}
 
  \begin{minipage}{0.19\hsize}
  \begin{center}
   \includegraphics[width=30mm]{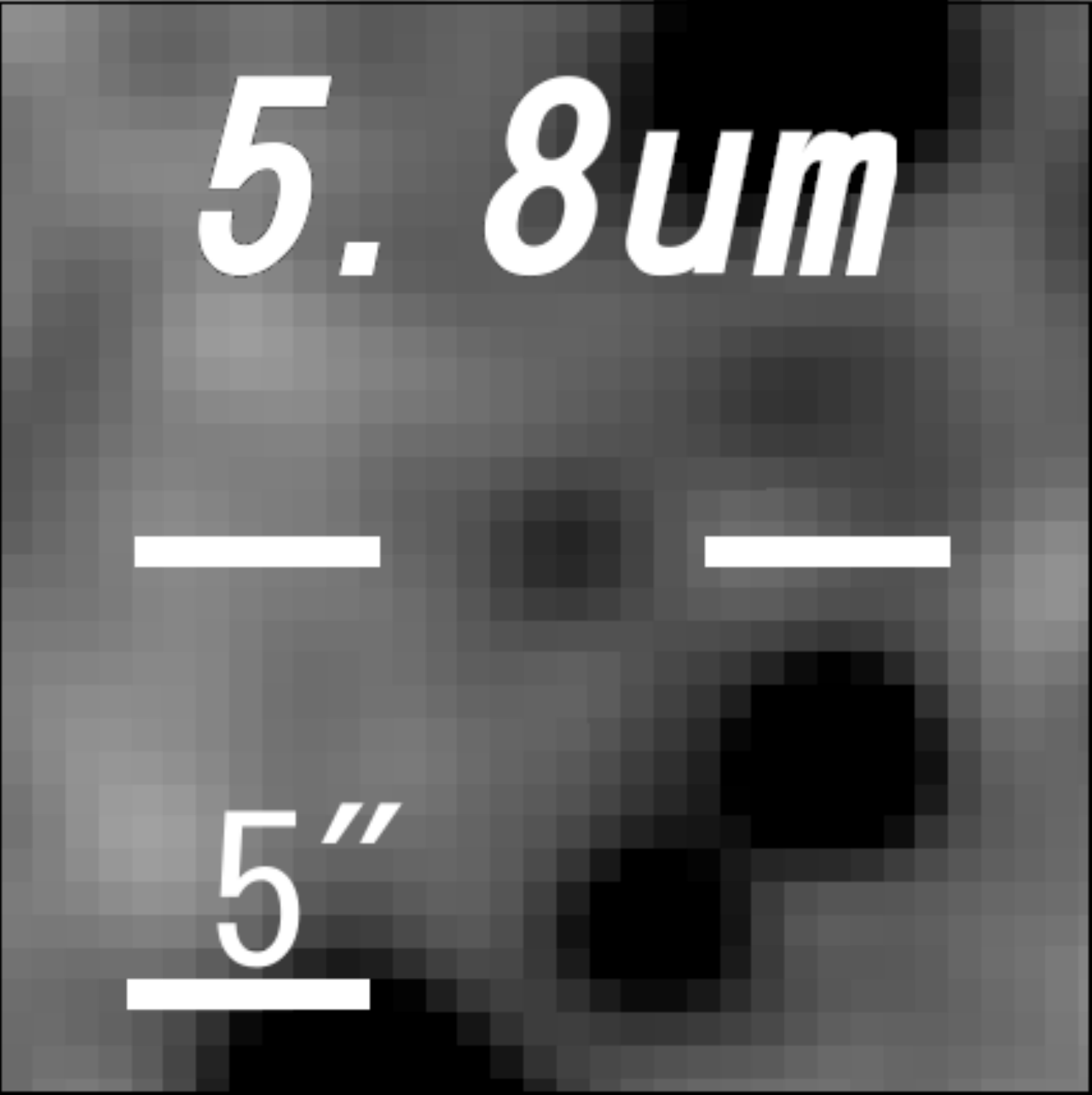}
  \end{center}
  \label{fig:one}
 \end{minipage}
 \begin{minipage}{0.19\hsize}
 \begin{center}
  \includegraphics[width=30mm]{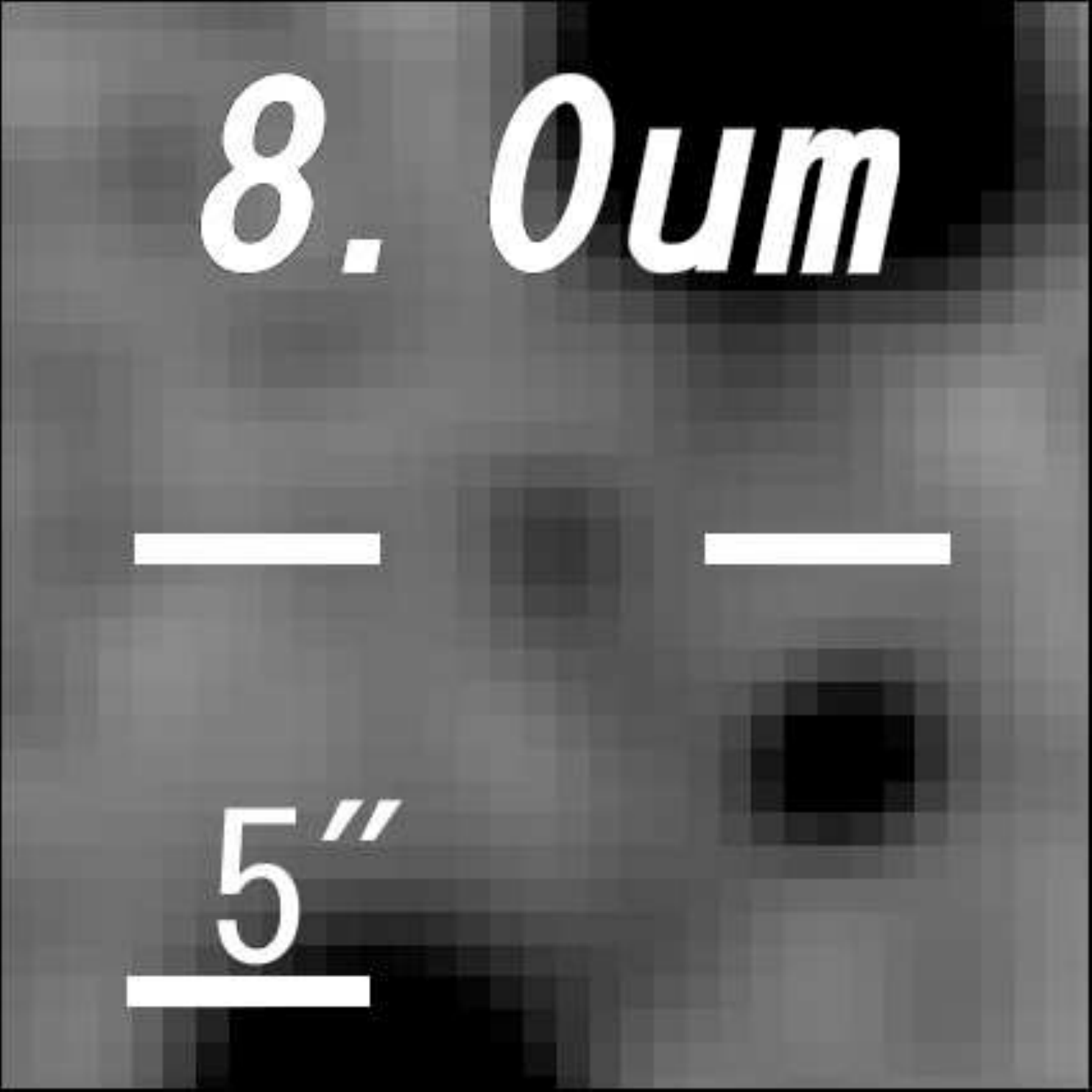}
 \end{center}
  \label{fig:two}
 \end{minipage}
 \begin{minipage}{0.19\hsize}
 \begin{center}
  \includegraphics[width=30mm]{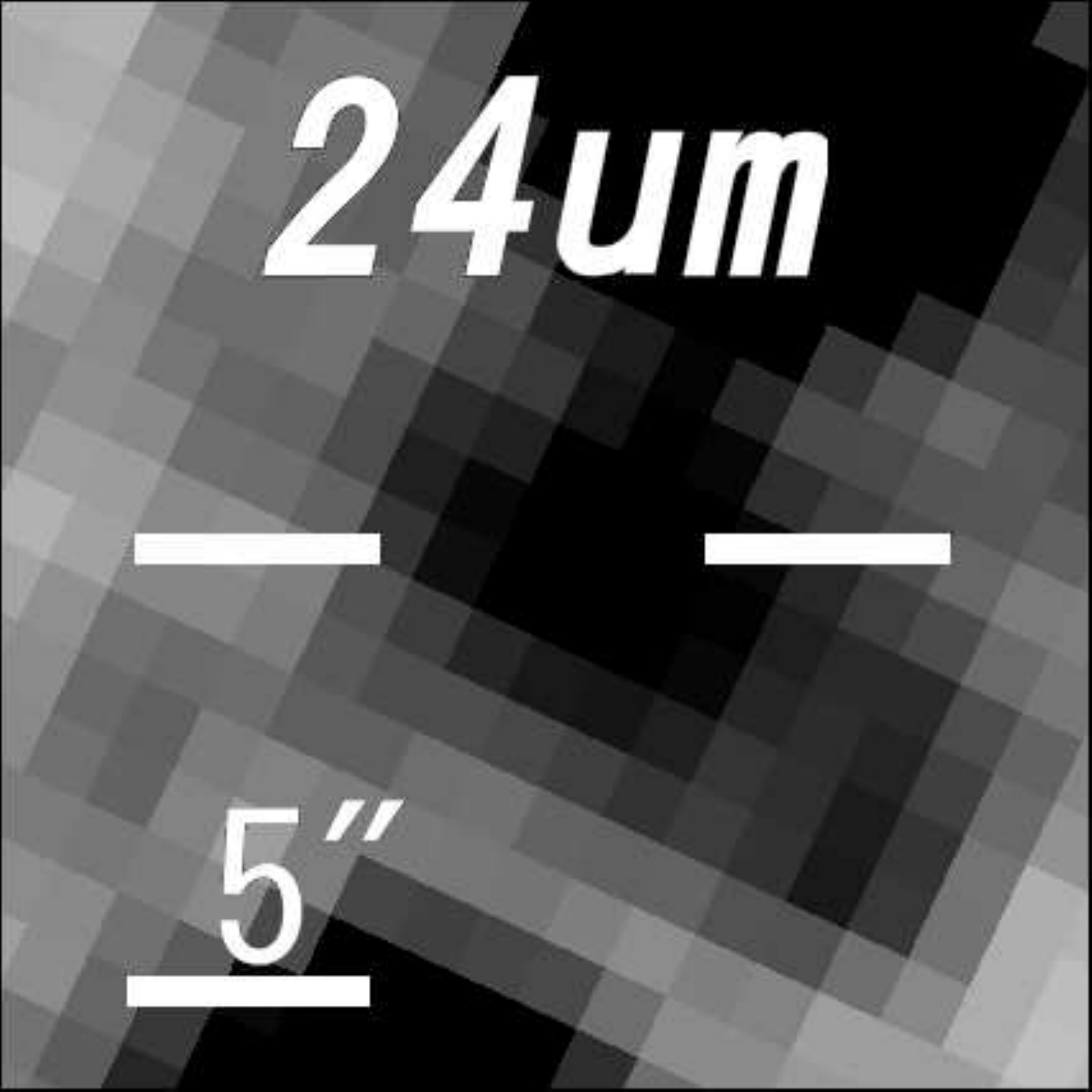}
 \end{center}
  \label{fig:three}
 \end{minipage}
  \begin{minipage}{0.19\hsize}
 \begin{center}
  \includegraphics[width=30mm]{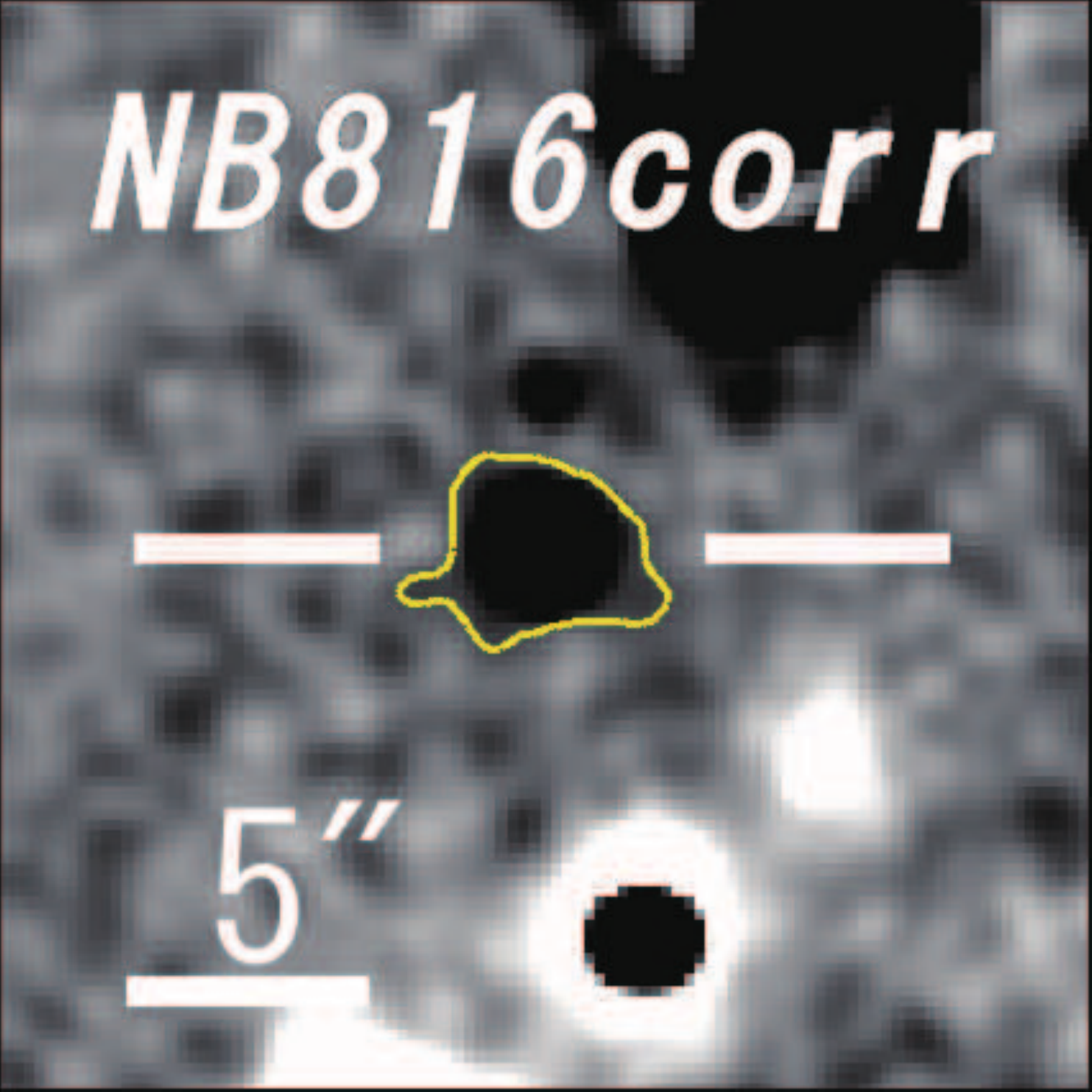}
 \end{center}
  \label{fig:three}
 \end{minipage}
  \begin{minipage}{0.19\hsize}
 \begin{center}
  \includegraphics[width=30mm]{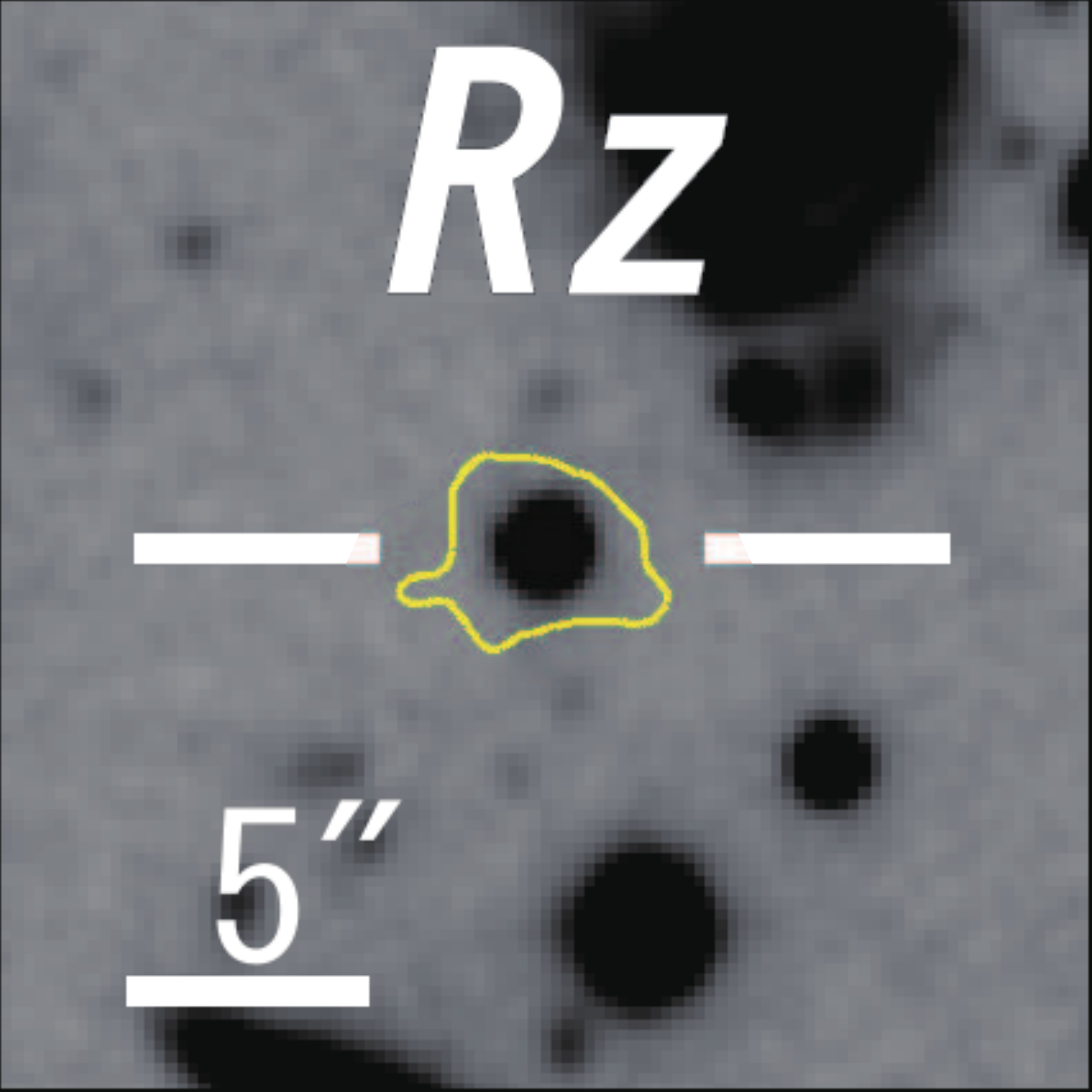}
 \end{center}
  \label{fig:three}
 \end{minipage}
 \caption{Optical to mid-infrared images of \ob. We display $20\arcsec\times20\arcsec$ images at {\it{BVRizJHK}}, {\it{Spitzer}}/IRAC $3.6\ \m{\mu m}$, $4.5\ \m{\mu m}$, $5.8\ \m{\mu m}$, $8.0\ \m{\mu m}$, and  {\it{Spitzer}}/MIPS $24\ \m{\mu m}$ bands. We also show ${NB816_\m{corr}}$ and ${\it Rz}$ images taken from Y13, which correspond to the {\sc [Oii]} and continuum emission images, respectively. The $5\arcsec$ scale bar is plotted at the bottom left in each image. The yellow contours in the ${\it{NB816_\m{corr}}}$ and ${\it Rz}$ images show the isophotal area above $2\sigma\ \m{arcsec^{-2}}$ in the ${NB816_\m{corr}}$ image as explained in Section \ref{tsso}.\label{fig_snap}}
\end{figure*}

Our target, {\ob}, is covered by deep optical and infrared imaging data 
from several ground and space-based surveys: 
the Subaru \textit{XMM-Newton} Deep Survey \citep[SXDS;][]{2008ApJS..176....1F}, 
the UKIDSS Ultra Deep Survey \citep[UDS;][]{2007MNRAS.379.1599L,2007MNRAS.375..213W}, 
and the \textit{Spitzer} public legacy survey of the UKIDSS UDS (SpUDS; PI: J. Dunlop).  
Figure \ref{fig_snap} shows the multi-wavelength images of {\ob}: 
the $BVRiz$ images from the SXDS, 
the $JHK$ images from the UKIDSS UDS, 
and the IRAC Ch1 ($3.6\ \mu$m), Ch2 ($4.5\ \mu$m), Ch3 ($5.8\ \mu$m), Ch4 ($8.0\ \mu$m) and MIPS $24\ \mu$m images from the SpUDS. 
We also present the $NB816_{\m{corr}}$ and $Rz$ images produced by Y13. 
Although the \textit{Herschel} SPIRE data are taken by the \textit{Herschel} Multi-tiered Extragalactic Survey \citep[HerMES;][]{2012MNRAS.424.1614O} in this field, 
the image of {\ob} is severely affected by source confusion.  
Hence, we do not use the SPIRE data.

The magnitudes in the $BVRiz$, $JHK$, $3.6\ \mu$m, and $4.5\ \mu$m bands are taken from Y13.
In this study, 
we estimate the magnitudes of {\ob} 
in the $5.8\ \m{\mu m}$, $8.0\ \m{\mu m}$, and $24\ \m{\mu m}$ bands. 
We obtain the magnitudes of the $5.8\ \mu$m and $8.0\ \mu$m bands with \texttt{MAG{\_}AUTO} of SExtractor \citep{1996A&AS..117..393B}.
The magnitude of the $24\ \mu$m band is estimated from {\sc galfit} \citep{2002AJ....124..266P} modeling, 
since \ob\ is significantly blended 
with the three neighboring objects.
We fit four objects, including \ob, with PSF profiles whose flux amplitudes and positions are free parameters.
Table \ref{tab_mag} summarizes the magnitudes of {\ob} estimated from the multi-wavelength imaging data. 

\begin{deluxetable}{lc}
\tablecaption{Magnitudes of \ob\label{tab_mag}}
\tablewidth{0pt}
\tablehead{
\colhead{Band} & \colhead{Magunitude}
}
\startdata
$B$ & $23.78\pm0.02$\\
$V$ & $23.42\pm0.02$\\
$R$ & $23.23\pm0.02$\\
$i$ & $22.90\pm0.02$\\
$z$ & $22.44\pm0.03$\\
$J$ & $22.05\pm0.04$\\
$H$ & $21.76\pm0.04$\\
$K$ & $21.39\pm0.03$\\
$m(3.6\m{\mu m})$ & $20.98\pm0.02$\\
$m(4.5\m{\mu m})$ & $21.09\pm0.03$\\
$m(5.8\m{\mu m})$\tablenotemark{a} & $21.6\pm0.3$\\
$m(8.0\m{\mu m})$\tablenotemark{a} & $21.9\pm0.4$\\
$m(24\m{\mu m})$ & $19.0\pm0.7$
\enddata
\tablenotetext{a}{$m(5.8\m{\mu m})$ and $m(8.0\m{\mu m})$ are marginally detected at the $2-3$ and $1-2$ $\sigma$ levels, respectively.}
\end{deluxetable} 
%

%

Y13 have derived the stellar population properties of {\ob}  
by fitting model spectral energy distributions (SEDs) with the observed SED (see Figure 3 of Y13).
The model SEDs have been constructed with the \citet{2003MNRAS.344.1000B} stellar population synthesis code. 
Y13 have assumed a constant star formation history, 
the \citet{1955ApJ...121..161S} initial mass function (IMF) 
with lower and upper cutoff masses of 0.1 and 100 \Msun, 
the dust attenuation law of \citet{2000ApJ...533..682C}, 
and the solar metallicity. 
The SED fitting results are presented in Table \ref{tab_sed}. 
The color excess is $E(B-V)=0.31\pm0.02$, implying that \ob\ is moderately dusty.
{The stellar age and the stellar mass are $270\pm80\ \m{Myr}$ and $(1.6\pm0.2)\times10^{10}\ M_{\odot}$, respectively.
The SFR is ${ \m{SFR_{SED}}}=72\pm8\ M_{\odot}\ \m{yr^{-1}}$.
Although we derive SFRs using different SFR indicators in Section \ref{sb}, ${\m{SFR_{SED}}}$ is our best estimate of \ob's SFR.
This is because ${\m{SFR_{SED}}}$ does not strongly depend on the metallicity but includes the extinction correction.}
The specific star formation rate (SSFR) is estimated to be high, $(45\pm8)\times10^{-10}\ \m{yr}^{-1}$, suggesting that \ob\ is a starburst galaxy.

\begin{deluxetable}{lc}
\tablecaption{Stellar Population Properties of \ob\label{tab_sed}}
\tablewidth{0pt}
\tablehead{
\colhead{Quantity} & \colhead{Value}
}
\startdata
$E(B-V)$ & $0.31\pm0.02\ \m{mag}$\\
Stellar Age & $270\pm80\ \m{Myr}$\\
Stellar Mass & $1.6\pm0.2\ \times10^{10}\ M_{\odot}$\\
${\m{SFR_{SED}}}$ \tablenotemark{a}  & $72\pm8\ M_{\odot}\ \m{yr}^{-1}$\\
${\m{SFR}_{\textsc{[Oii]}}}$\tablenotemark{a}  & $81\pm27\ M_{\odot}\ \m{yr}^{-1}$\\
${\m{SFR_{H\beta}}}$\tablenotemark{a}  & $96\pm20\ M_{\odot}\ \m{yr}^{-1}$\\
${\m{SFR_{24\ {{\mu}m}}}}$\tablenotemark{a}  & $100\pm70\ M_{\odot}\ \m{yr}^{-1}$
\enddata
\tablenotetext{a}{$\m{SFR_{SED}}$, ${\m{SFR}_{\textsc{[OII]}}}$, $\m{SFR_{H\beta}}$, and $\m{SFR_{24\ {{\mu}m}}}$ are SFRs derived from the SED fitting, the [O{\sc ii}] luminosity, the {H$\beta$} luminosity, and the {24\ {{$\mu$}m}} flux, respectively.}
\end{deluxetable}

\section{Observations and Data Analyses}\label{obdr}

\subsection{MOSFIRE}\label{obda_mos}


We observed {\ob} with the MOSFIRE instrument \citep{2012SPIE.8446E..0JM} 
on the Keck-I telescope on 2013 October 8 (PI: M. Ouchi). 
A spectrum of \ob\ was obtained as a mask filler of this observation.
We carried out $Y$-band spectroscopy,
which covered the wavelength range of $9800$--$11000$ {\AA}. 
{Thus this observation targeted H$\beta$ and  {\sc [Oiii]}$\lambda\lambda4959,5007$ lines redshifted to $z\sim1.2$.}
The slit width was 0.\carcsec7. 
We used an ABAB dither pattern with individual exposures of 180 seconds. 
The total integration time was 2.4 hours, 
and the average seeing size was 0.\carcsec9 in an FWHM. 
The pixel scale was 0.\carcsec18 pixel$^{-1}$, 
and the spectral resolution was $R=3388$. 
We took the spectrum of a standard star, \objectname{HIP 13917}, for our flux calibration.

We reduce the data 
using the MOSFIRE data reduction pipeline.\footnote{http://code.google.com/p/mosfire} 
This pipeline performs flat fielding, wavelength calibration, sky subtraction, 
and cosmic ray removal 
before producing a combined two-dimensional spectrum for each slit. 
We then extract one-dimensional spectra from the two-dimensional reduced spectra 
using the {\sc iraf} task {\tt apall}. 
We sum the fluxes of {3.\carcsec24 ($18$ pixels)} at each wavelength bin that covers {about 95\%} of the emission.

To determine total line fluxes, we calculate a slit loss. 
Firstly, a radial profile of each line is estimated with the spatial distribution in the two-dimensional spectrum. 
We assume that the radial profile is identical in all directions, 
and calculate a correction factor to be 1.2.
The $3\sigma$ limiting flux density is 
$\simeq 5\times10^{-18}\ \m{erg\ s^{-1}\ cm^{-2}\ \AA^{-1}}$ 
in the wavelength range of $9800-11000$ {\AA}.

%

\subsection{LDSS3}


We conducted deep spectroscopic observations for {\ob} 
using the LDSS3 
on the 6.5 m Magellan II (Clay) telescope 
on 2013 November 3 (PI: M. Rauch).  
We adopted a $0 \farcs 8 \times 3 \farcs 5$ slitlet, 
and used the volume phase holographic (VPH)-red grism with the sloan-$i$ filter, 
which provided a spectral coverage between $6700$ and $8600$ {\AA}. 
{{\sc [Oii]} lines redshifted to $z\sim1.2$ were targeted in this configuration.}
In addition, 
we took spectra with the VPH-blue grism and the w4800--7800 filter, 
which covered the wavelength range from $4800$ {\AA} to $6600$ {\AA}. 
{This configuration targeted metal absorption lines such as Mg{\sc ii} and Fe{\sc ii}.}
The on-source exposure times 
with the VPH-red and VPH-blue configurations 
were $0.5$ hours ($= 1 \times 1800$ seconds) 
and $2.5$ hours ($= 3 \times 3000$ seconds), respectively. 
The average seeing size was $0 \farcs 6$ in FWHM. 
The VPH-red (blue) grism had a spectral resolution of $R\sim1710\ (1800)$. 
The LDSS3 instrument had a pixel scale of $0 \farcs 189$ pixel$^{-1}$. 
\objectname{Feige 110} was observed for our flux calibration.

The LDSS3 data are reduced 
by the Carnegie Observatories reduction package, 
{\sc cosmos}.\footnote{http://code.obs.carnegiescience.edu/cosmos} 
After bias subtraction, flat fielding, wavelength calibration, and rectification, 
one-dimensional spectra are extracted in the same manner as Section \ref{obda_mos}, 
except that the spectra are summed over {2.\carcsec27 ($12$ pixels)} 
along the spatial axis {to cover about 95\% of the emission}. 
We estimate a slit-loss correction factor to be 1.3 by the procedure same as Section \ref{obda_mos}.
The $3 \sigma$ flux density limit of the VPH-red spectrum is calculated to be 
$\simeq 8\times10^{-18}\ \m{erg\ s^{-1}\ cm^{-2}\ \AA^{-1}}$ 
in the wavelength range of $6700$--$8600$ {\AA}\footnote{The VPH-blue grism spectrum is not flux-calibrated, because we only need velocity centroids, EWs, and FWHM line widths in this study.}.

\begin{figure*}
  \begin{minipage}{0.32\hsize}
  \begin{center}
   \includegraphics[width=60mm,bb=15 5 435 380,clip]{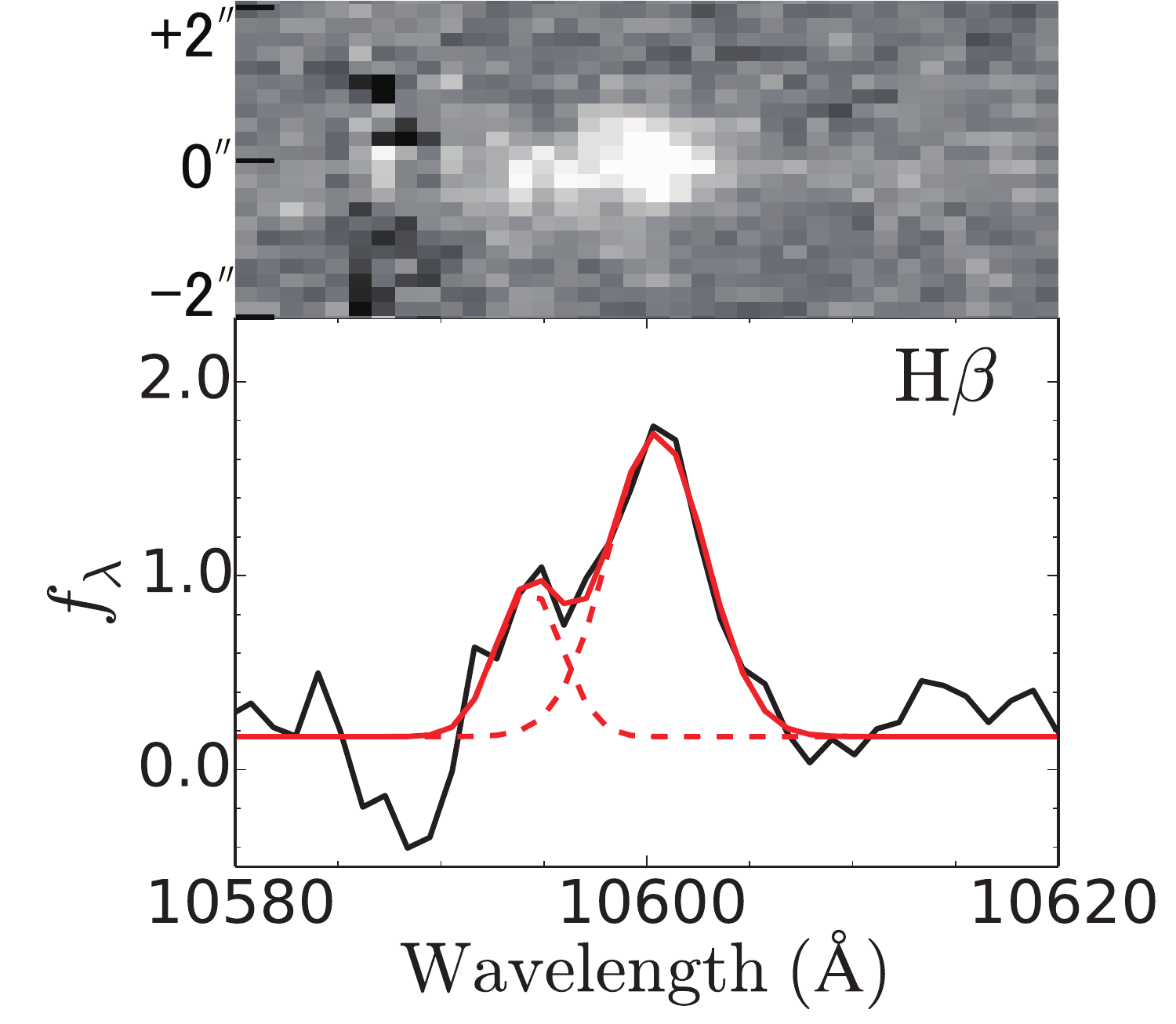}
  \end{center}
  \label{fig:one}
 \end{minipage}
 \begin{minipage}{0.32\hsize}
 \begin{center}
  \includegraphics[width=60mm,bb=15 5 435 380,clip]{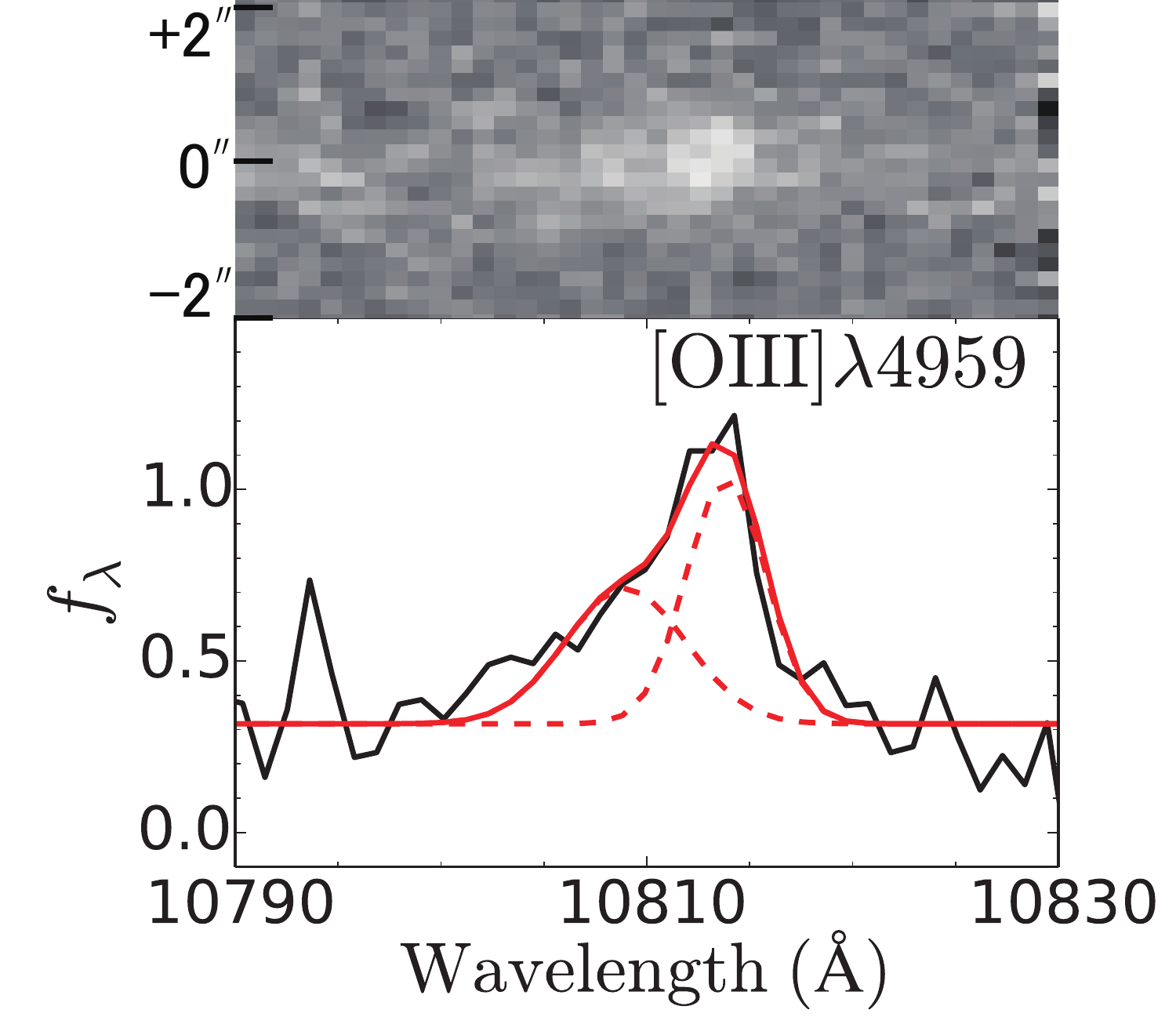}
 \end{center}
  \label{fig:two}
 \end{minipage}
 \begin{minipage}{0.32\hsize}
 \begin{center}
  \includegraphics[width=60mm,bb=15 5 435 380,clip]{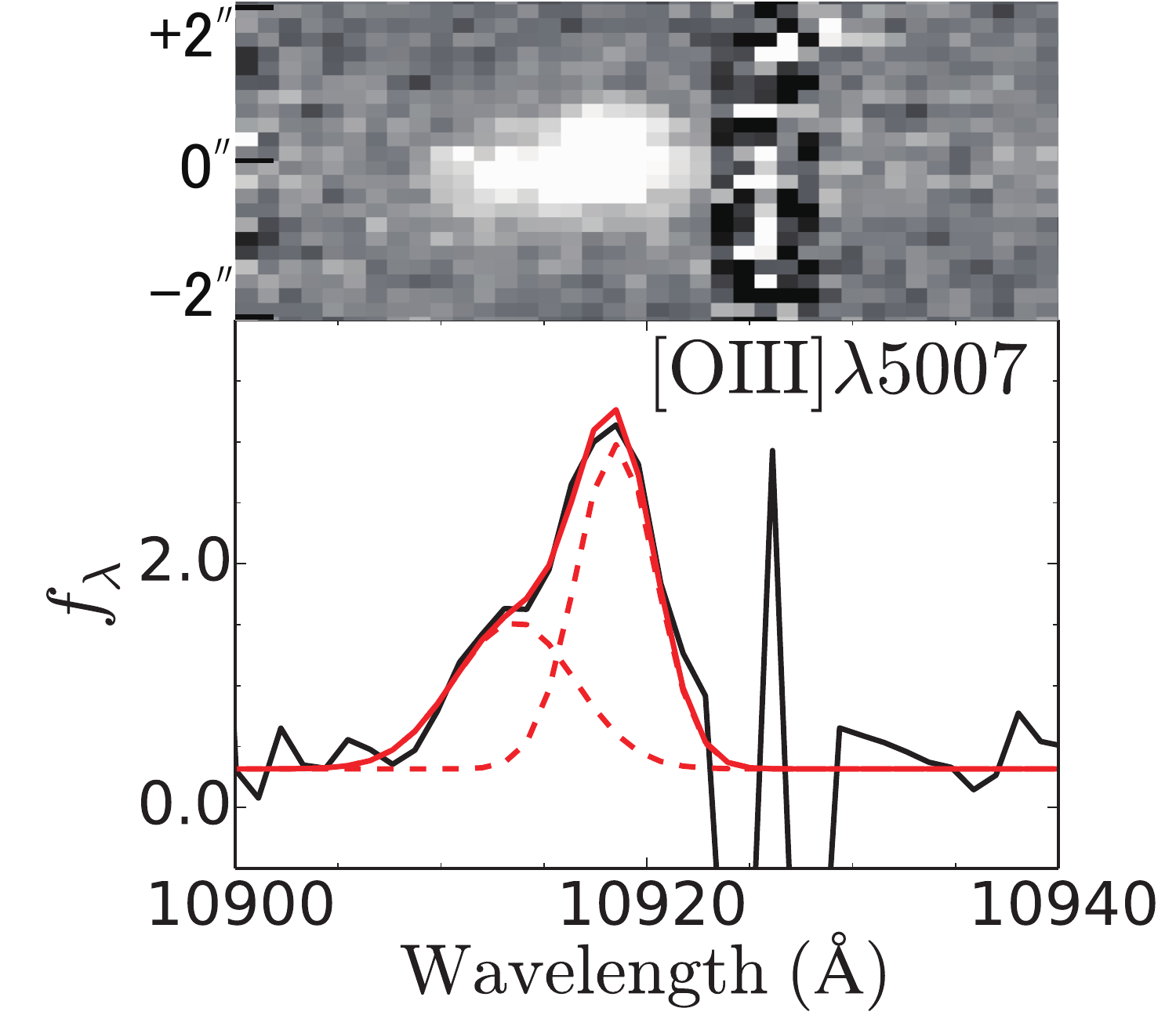}
 \end{center}
  \label{fig:three}
 \end{minipage}
 \caption{MOSFIRE spectra of \ob\ at the wavelengths of H$\beta$ and {\sc [Oiii]}$\lambda\lambda4959,5007$ emission lines in the observed frame. The top panels present the two-dimensional spectra. In the bottom panels, the black solid lines indicate the observed spectra, and all emission lines show asymmetric profiles. The red solid lines denote the best-fit functions, and the red dashed lines represent Gaussian components of the best-fit functions. The units of the vertical axes are $10^{-17}\ \m{erg\ s^{-1}\ cm^{-2}\ \AA^{-1}}$. \label{fig_mos}}
\end{figure*}
\begin{center}
\begin{deluxetable*}{lcccccccccc}
\setlength{\tabcolsep}{0.1cm}
\tablecolumns{11}
\tablecaption{Emission Line Properties of \ob.\label{tab_el}}
\tablewidth{0pt}
\tablehead{
\colhead{ Line}& \colhead{ $z_\m{B}\ /\ z_\m{R}$}   & \colhead{ $f_{\m{total}}^{\m{corr}}$} & \colhead{ $f_{\m{total}}^{\m{obs}}$}  & \colhead{ $f_{\m{B}}^{\m{obs}}$} & \colhead{ $f_{\m{R}}^{\m{obs}}$}  & \colhead{ $\m{FWHM}_{\m{B}}\ /\ \m{FWHM}_{\m{R}}$} \\
& \colhead{ (1)} & \colhead{ (2)}& \colhead{ (3)}& \colhead{ (4)}& \colhead{ (5)}& \colhead{ (6)} 
}
\startdata
 {\sc [Oii]}$\lambda3726$ &  $1.1800\pm0.0007$\tablenotemark{a} &  $35.3\pm11.6$ &  $6.5\pm1.8$ &  \dots &  \ldots &   $120\pm60$\tablenotemark{a} 
\\
 {\sc [Oii]}$\lambda3729$ &  $1.1800\pm0.0007$\tablenotemark{a} &  $37.5\pm11.6$ &  $6.9\pm1.8$ &  \dots &  \ldots &   $120\pm60$\tablenotemark{a} 
\\
 $\m{H\beta}$ &  $1.1794\pm0.0004$\ /\ $1.1807\pm0.0007$&  $53.2\pm12.3$ &  $14.3\pm2.6$ &  $3.5\pm1.8$ &  $10.8\pm1.9$ &  $70\pm50$\ /\ $130\pm50$ 
\\
 {\sc [Oiii]}$\lambda4959$ &  $1.1797\pm0.0007$\ /\ $1.1806\pm0.0005$ &  $26.9\pm6.8$ &  $7.4\pm1.5$ &  $3.4\pm1.0$ &  $4.1\pm1.0$ & $120\pm100$\ /\ $100\pm50$
\\
 {\sc [Oiii]}$\lambda5007$ & $1.1797\pm0.0007$\ /\ $1.1806\pm0.0005$ &  $90.2\pm21.2$ &  $25.2\pm4.5$ &  $10.3\pm3.1$ &  $15.0\pm3.1$ &  $120\pm100$\ /\ $100\pm50$ 
\enddata
\tablecomments{All fluxes and FWHM line widths are in units of $\m{10^{-17}\ erg\ s^{-1} cm^{-2}}$ and $\kms$, respectively. {All FWHM line widths are corrected for the instrumental broadening.} Columns: {(1) Redshifts for the blue ($z_\m{B}$) and red ($z_\m{R}$) components.} (2) Total flux of two components after the slit-loss and dust-extinction corrections. (3) Total flux of the two components corrected only for the slit loss. (4)(5) Fluxes of blue $(f_{\m{B}}^{\m{obs}})$ and red $(f_{\m{R}}^{\m{obs}})$ components after the slit-loss correction. (6) FWHM line widths of the blue $(\m{FWHM}_{\m{B}})$ and red $(\m{FWHM}_{\m{R}})$ components.}
\tablenotetext{a}{The blue and red components of the {\sc [Oii]} lines are not resolved.}
\end{deluxetable*}
\end{center}
\section{Results}\label{a}
\subsection{MOSFIRE}\label{res_mos}

Figure \ref{fig_mos} shows our MOSFIRE spectra at the wavelengths of {\sc [Oiii]} and H$\beta$.
We detect strong {\sc [Oiii]} and H$\beta$ emission lines.
These lines appear to have two components.
We hereafter call the shorter-wavelength component "the blue component", and the longer-wavelength component "the red component".
{The red components appear to be spatially more extended than the blue components.
The implications of these two-component profiles and the spatial extents are discussed in Section \ref{what}.}
The two-component H$\beta$ line is fitted with two Gaussian functions. 
The {\sc [Oiii]} lines are the doublet lines, and each of the doublet lines has the blue and red components.
Hence, we fit the {\sc [Oiii]} lines with four Gaussian functions, assuming that each component of the {\sc [Oiii]} doublet lines has the same line widths.

The results of the spectral line fittings are presented in Table \ref{tab_el}.
The redshift of the blue (red) component of {\sc [Oiii]} lines is in agreement with that of the blue (red) component of H$\beta$ line within the $1\sigma$ uncertainties.
We derive the average redshifts of the blue and red components to be $\left< z_\m{B} \right> =1.1795\pm0.0003$ and $\left< z_\m{R} \right> =1.1806\pm0.0004$, respectively, from the H$\beta$ and {\sc [Oiii]} lines.
The velocity difference derived from $\left< z_\m{B} \right>$ and $\left< z_\m{R} \right>$ is $170\pm50\ \kms$.
The average redshift of all components is $z_{\m{sys}}=1.1800\pm0.0002$, and we define this average value as the systemic redshift of \ob.

{The estimated FWHM line widths are also summarized in Table \ref{tab_el}.}
These FWHM line widths are corrected for the instrumental broadening on the assumption that their intrinsic profiles are Gaussian.
The FWHM line widths of the {\sc [Oiii]} lines are comparable to those of the H$\beta$ line within the $1\sigma$ uncertainties.
The average FWHM line widths of blue and red components are $\left< \m{FWHM_B} \right> =90\pm50\ \kms$ and $\left< \m{FWHM_R} \right> =120\pm40\ \kms$, respectively.



We correct the fluxes for reddening by adopting the \citet{2000ApJ...533..682C} extinction law and $E(B-V)=0.31$ mag presented in Table \ref{tab_sed}. 
The fluxes after the dust-extinction correction are shown in Table \ref{tab_el}.

\subsection{LDSS3}\label{ss_ldss3}

\begin{figure}
\begin{center}
\includegraphics[width=60mm,bb=10 5 422 395,clip]{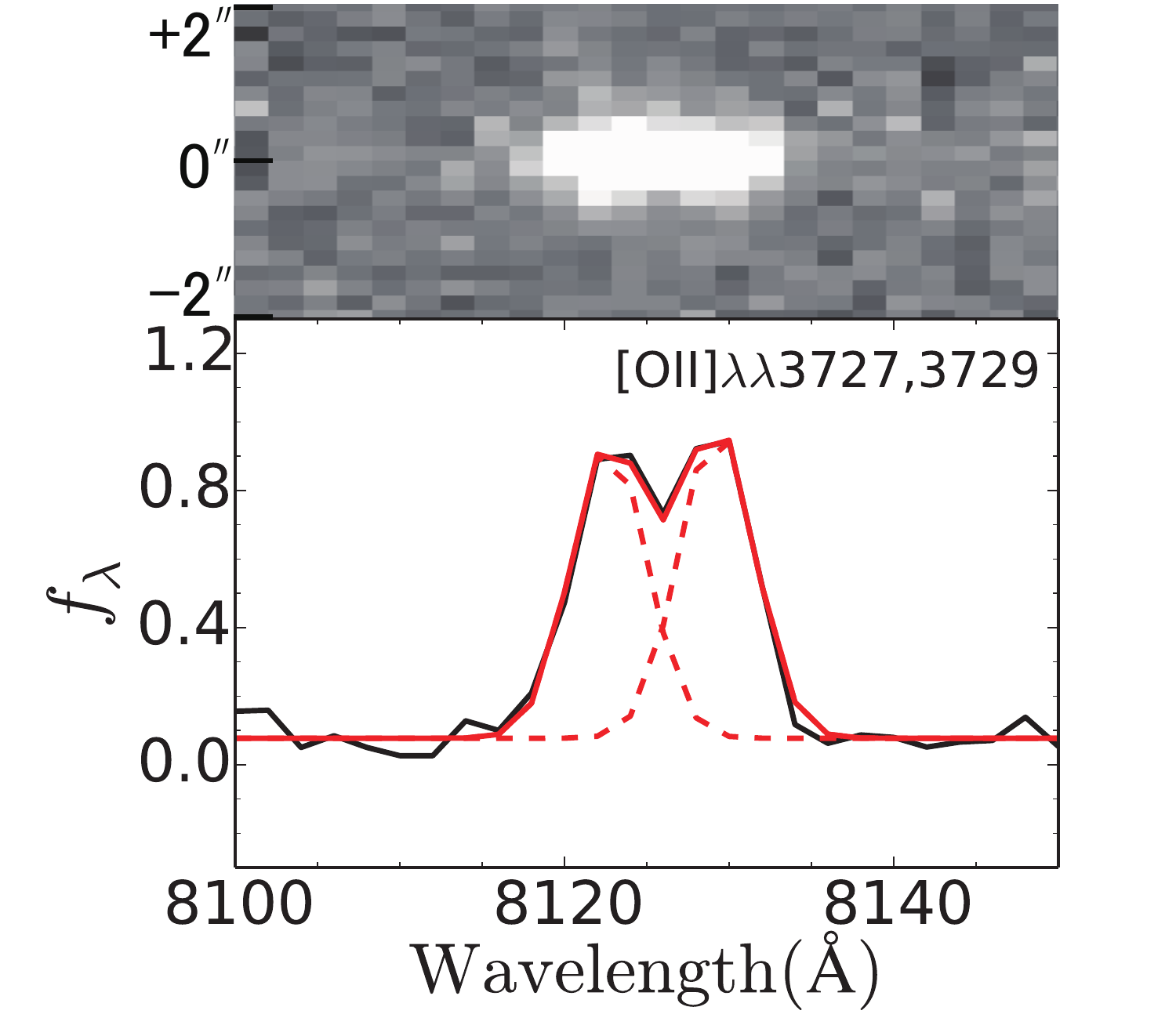}
\end{center}
\caption{LDSS3 spectrum of \ob\ taken with the VPH-red grism. Strong {\sc [Oii]}$\lambda\lambda3726,3729$ emission lines are detected. The top panel shows the two-dimensional spectrum. In the bottom panel, the black solid line shows the observed spectrum, and each of the doublet lines has a symmetric profile. The red solid line indicates the best-fit double Gaussian function, and the red dashed line represents each Gaussian component. The unit of the vertical axis is $10^{-17}\ \m{erg\ s^{-1}\ cm^{-2}\ \AA^{-1}}$. \label{fig_ldss3_red}}
\end{figure}

Figure \ref{fig_ldss3_red} presents our LDSS3 VPH-red grism spectrum at the wavelength of {\sc [Oii]}.
We identify {\sc [Oii]} emission lines at the high significance level.
In contrast  to the MOSFIRE spectra of the H$\beta$ and [{\sc Oiii}] lines, the LDSS3 spectrum of the 
{\sc [Oii]} line does not show two components but one component.
The non-detection of the blue and red components of the {\sc [Oii]} doublet lines is due to the low resolution of the LDSS3 VPH-red data.
The resolution of the LDSS3 VPH-red data ($R\sim1710$) is lower than that of the MOSFIRE data ($R=3388$), and the velocity difference of $170\pm50\ \kms$ (Section \ref{res_mos}) is not resolved in the LDSS3 data. 
Thus, each of the doublet lines has a one-component profile, and we fit {\sc [Oii]} doublet lines with two Gaussian functions.

Table \ref{tab_el} summarizes the fitting results. 
The redshift of {\sc [Oii]} doublet lines is consistent with the systemic redshift, $z_{\m{sys}}=1.1800\pm0.0002$. 
{The {\sc [Oii]} flux corrected for the slit loss is $f_{\m{total}}^{\m{obs}}$({\sc{[Oii]}})$=(13.4\pm5.1)\times10^{-17}\ \m{erg\ s^{-1}\ cm^{-2}\ \AA^{-1}}$, which is consistent with the {\sc [Oii]} flux derived from the $NB816_\m{corr}$ image, $14\times10^{-17}\ \m{erg\ s^{-1}\ cm^{-2}\ \AA^{-1}}$.}
The fluxes are corrected for the dust extinction in the same manner as Section \ref{res_mos}.

\begin{figure*}
 \begin{minipage}{0.48\hsize}
 \begin{center}
  \includegraphics[width=80mm,bb=0 0 430 311,clip]{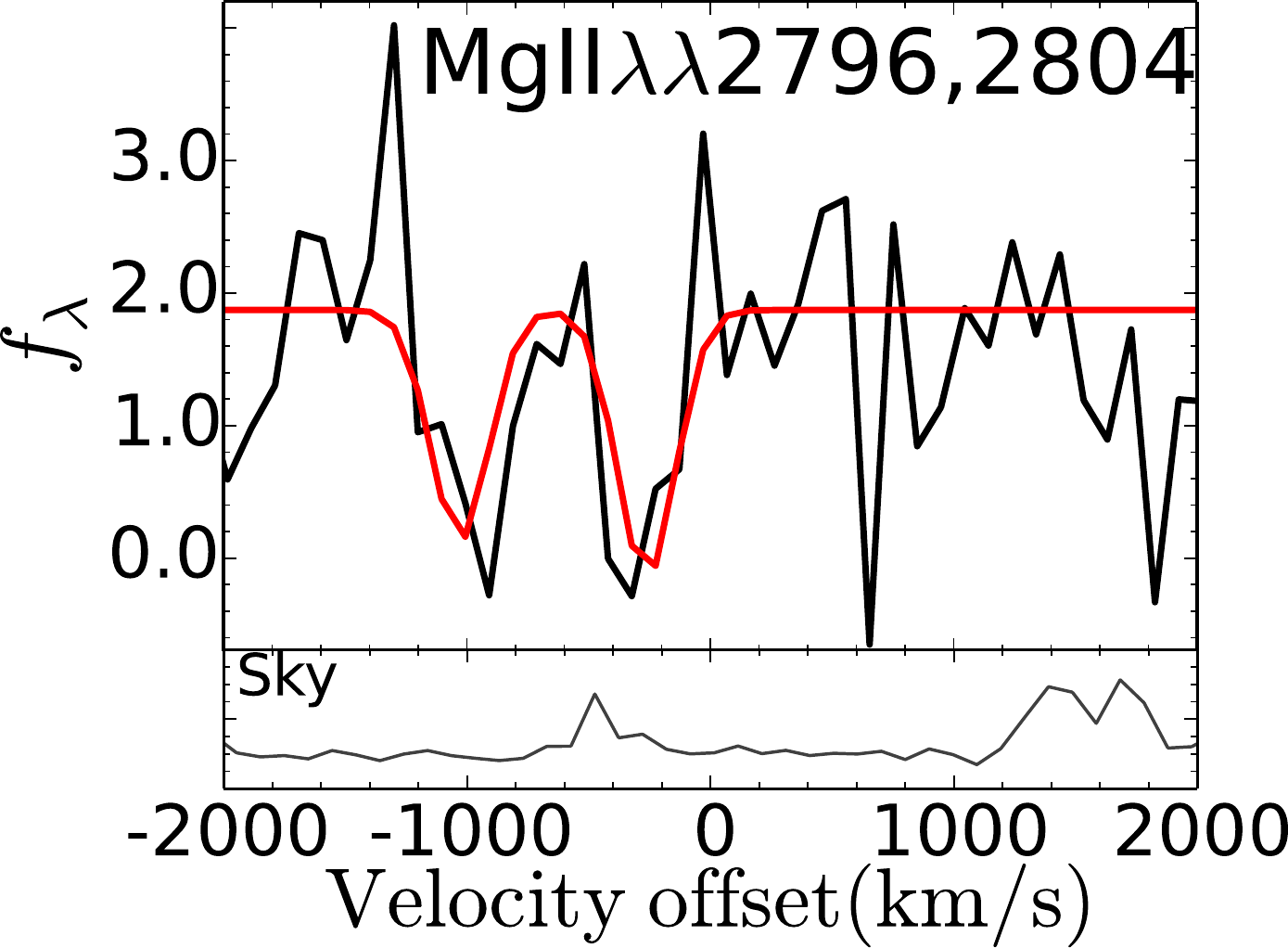}
 \end{center}
  \label{fig:two}
 \end{minipage}
 \begin{minipage}{0.48\hsize}
 \begin{center}
  \includegraphics[width=80mm,bb=0 0 430 311,clip]{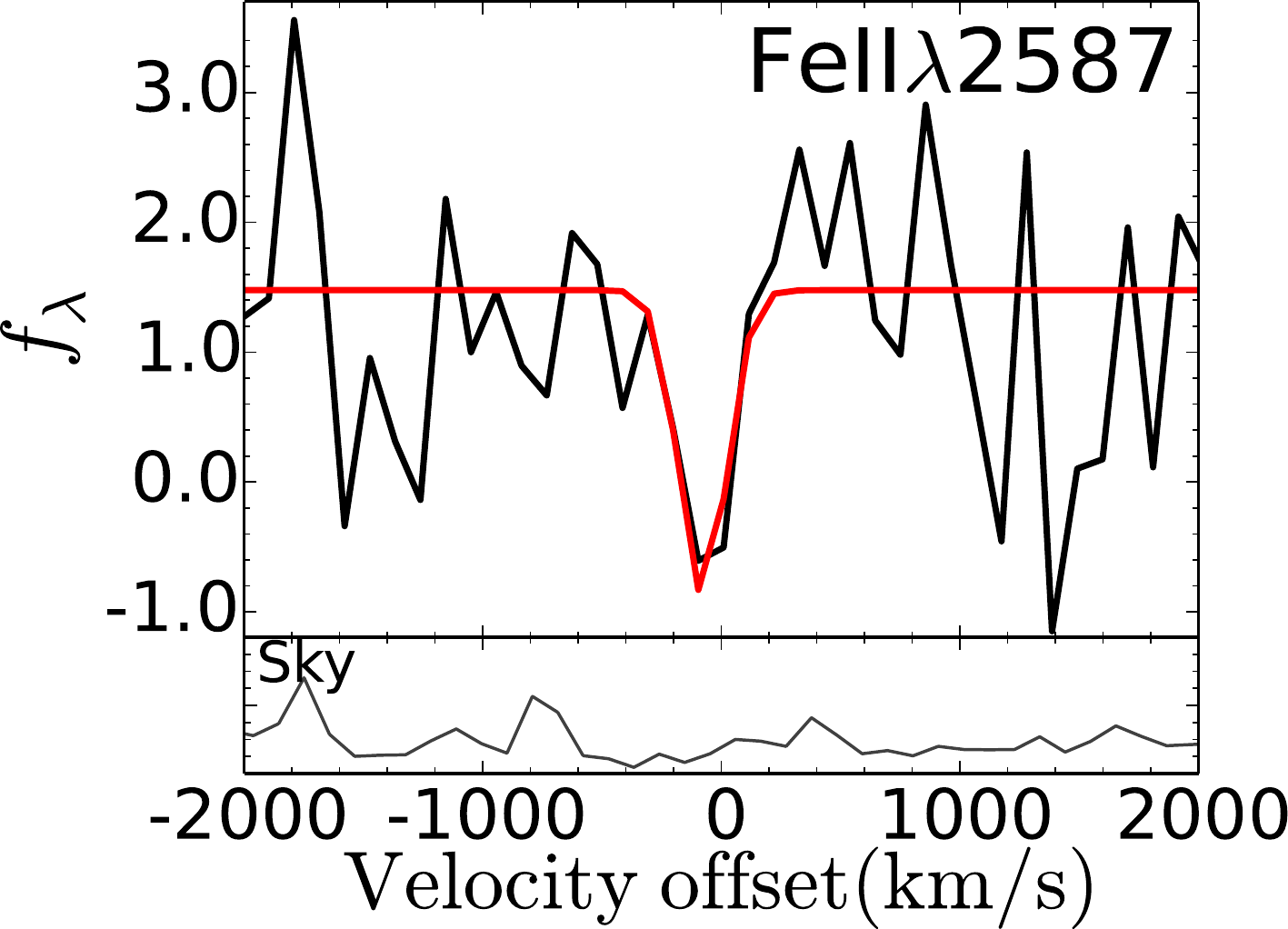}
 \end{center}
  \label{fig:three}
 \end{minipage}
 \caption{LDSS3 spectra of \ob\ taken with the VPH-blue grism. The blueshifted Mg{\sc ii}$\lambda\lambda$2796,2804 (left panel) and Fe{\sc ii}$\lambda$2587 (right panel) absorption lines are identified at the 5.5 and 2.7 $\sigma$ levels, respectively. In the top panels, black lines represent spectra of \ob, and red lines denote the best-fit functions. The velocity offsets of Mg{\sc ii}$\lambda\lambda$2796,2804 and Fe{\sc ii}$\lambda$2587 lines from the systemic velocity
are $\Delta{v}=-260 \pm 40$ km s$^{-1}$
and $\Delta{v}=-80 \pm 50$ km s$^{-1}$, respectively. The bottom panels show the background sky spectra. The units of the vertical axes are arbitrary. \label{fig_ldss3_blue}}
\end{figure*}

Figure \ref{fig_ldss3_blue} indicates our LDSS3 VPH-blue grism spectra at the wavelengths of Mg{\sc ii}$\lambda\lambda$2796,2804 and Fe{\sc ii}$\lambda\lambda$2587,2600. 
We identify Mg{\sc ii}$\lambda\lambda$2796,2804 and Fe{\sc ii}$\lambda$2587 absorption lines at the 5.5 and 2.7 $\sigma$ levels, respectively. 
We do not examine the Fe{\sc ii}$\lambda$2600 absorption line, 
since the sky background at this wavelength is relatively high. 
The Mg{\sc ii}$\lambda\lambda$2796,2804 and Fe{\sc ii}$\lambda$2587 lines are fitted with two and one Gaussian function(s), respectively.

The fitting results are presented in Table \ref{tab_ldss3_blue}.
We find that the Mg{\sc ii}$\lambda\lambda$2796,2804 absorption lines are blueshifted from the systemic redshift by $|\Delta{v}|=260 \pm 40$ km s$^{-1}$.
The Fe{\sc ii}$\lambda$2587 absorption line is also blueshifted, but the velocity offset is $|\Delta{v}|=80 \pm 50$ km s$^{-1}$.
Although the Fe{\sc ii}$\lambda$2587 line is poorly identified, the velocity offset of the Fe{\sc ii}$\lambda$2587 line may not be the same as that of the Mg{\sc ii}$\lambda\lambda$2796,2804 lines.


\begin{deluxetable}{lccc}
\tablecaption{Absorption Line Properties of \ob.\label{tab_ldss3_blue}}
\tablewidth{0pt}
\tablehead{
\colhead{Line}& \colhead{$\m{EW}$} & \colhead{$\Delta{v}$} & \colhead{$\m{FWHM}$} \\
& \colhead{$\m{(\AA)}$} & \colhead{($\kms$)}& \colhead{($\kms$)}\\
& \colhead{(1)} & \colhead{(2)}& \colhead{(3)}
}
\startdata
Mg{\sc ii}$\lambda$2796 & $-2.6\pm0.7$ & $-260\pm40$ & $240\pm110$ \\
Mg{\sc ii}$\lambda$2803 & $-3.1\pm0.7$ & $-260\pm40$ & $240\pm110$ \\
Fe{\sc ii}$\lambda$2587 & $-3.0\pm1.1$ & $-80\pm50$ & $180\pm120$
\enddata
\tablecomments{Columns: (1) Rest-frame equivalent width. (2) Velocity offset from the systemic redshift. (3) FWHM line width {corrected for the instrumental broadening.}}
\end{deluxetable}

\section{Discussion}\label{r}



\subsection{Signature of Outflow\label{ao}}
In Section \ref{ss_ldss3}, we identify the Mg{\sc ii}$\lambda\lambda$2796,2804 and Fe{\sc ii}$\lambda$2587 absorption lines blueshifted by $|\Delta{v}|=260 \pm 40\ \m{\kms}$ and $|\Delta{v}|=80 \pm 50$ km s$^{-1}$, respectively. 
These blueshifted absorption lines are evidence of outflows, 
because blueshifted photons are absorbed by the outflowing gas along the sight of the stars.
From the blueshifted Mg{\sc ii} and Fe{\sc ii} absorption lines, the outflow velocities are estimated to be $260 \pm 40$ and $80 \pm 50$ km s$^{-1}$, respectively. 
The difference of these outflow velocities may be significant; the implications of this difference are discussed in Section \ref{diff}. 
\citet{2005ApJ...632..751R,2005ApJS..160...87R} suggest that outflow velocities of star-formation-driven (SF-driven) winds are $20-500 \ \m{\kms}$, while those of galaxies hosting AGNs are $70-600 \ \m{\kms}$.
Thus, the low outflow velocity of \ob, $80-260\ \kms$, is comparable to those of AGN-driven winds as well as those of the SF-driven winds.


In some galaxy spectra, the Mg{\sc ii} line exhibits a P Cygni profile, which is composed of a redshifted resonance emission line on the intrinsic absorption trough \citep[e.g.,][]{2009ApJ...703.1394M,2009ApJ...692..187W,2010ApJ...719.1503R,2011ApJ...728...55R,2011ApJ...734...24P,2011ApJ...743...46C,2012ApJ...759...26E,2012ApJ...760..127M,2013ApJ...770...41M}. 
{\citet{2011ApJ...728...55R} detect Mg{\sc ii} doublet emission in a starburst galaxy at $z=0.69$ whose rest-frame EW is $3.6\ \m{\AA}$.
Assuming this EW, we can exclude the presence of the Mg{\sc ii} doublet emission in our LDSS3 VPH-blue data at the $3.6\ \sigma$ level.}
\citet{2009ApJ...692..187W} suggest that the Mg{\sc ii} emission could be a weak AGN activity signature and/or backscattered light in the outflow. 
They report that blue and low-mass star-forming galaxies have Mg{\sc ii} emission lines stronger than red and high-mass star-forming galaxies. 
\citet{2012ApJ...760..127M} present the same trend, and suggest that this trend may be attributed to dust attenuation. 
Massive and red star-forming galaxies are dusty, and Mg{\sc ii} photons are absorbed by dust grains in the foreground gas.
Thus, the non-detection of the Mg{\sc ii} emission line in our spectrum implies that \ob\ would be dusty.

%

\subsection{Outflow Rate and Mass Loading Factor}
{Following the procedures of \citet{,2010ApJ...719.1503R},} we estimate the column density of the outflowing gas with the ratio of the Mg{\sc ii} doublet line EWs \citep[see also][]{1968dms..book.....S,1986ApJ...304..739J,2009ApJ...692..187W}. 
The EW ratio of the Mg{\sc ii} doublet lines, $\m{EW}_{2796}/\m{EW}_{2803}$, varies from 2 to 1 for optical depth, $\tau_0$, increasing from 0 to infinity. 
The Mg doublet EW ratio is {approximately} equal to $F(2\tau_0)/F(\tau_0)$, where $F(\tau_0)$ is given by
\begin{equation}
F(\tau_0)=\int_0^{\infty}(1-e^{-\tau_0\m{exp}(-x^2)})dx.
\end{equation}
After the doublet ratio is estimated, we numerically derive $\tau_0$. 
Once $\tau_0$ is given, we calculate the column density in atoms, $N(\m{Mg\textsc{ii}})$, using the equation from \citet{1968dms..book.....S}:
\begin{equation}
\begin{split}
\m{log}N(\m{Mg\textsc{ii}})=\m{log}\frac{|\m{EW}_{2803}|}{\lambda}-\m{log}\frac{2F(\tau_0)}{\pi^{1/2}\tau_0}-\m{log}\lambda{f_{2803}}\\-\m{log}C_f+20.053,
\label{eq_CD}
\end{split}
\end{equation}
where $\m{log}\lambda{f_{2803}}=2.933$ and $C_f$ is the covering fraction. 
Here we assume $C_f=1$.
The EW ratio, $\m{EW}_{2796}/\m{EW}_{2803}$, of \ob\ is $0.86\pm0.30$, whose best-estimate value is smaller than unity.
If we adopt the $1\sigma$ upper limit of the EW ratio, $\m{EW}_{2796}/\m{EW}_{2803}=1.16$, we obtain $\tau_0=5.9$, which corresponds to the $1\sigma$ lower limit. 
This $1\sigma$ lower limit yields $N(\m{Mg\textsc{ii}})>4.9\times10^{14}\ \m{cm^{-2}}$.
We assume $N(\m{Mg\textsc{ii}})=N(\m{Mg})$ and the abundance ratio of $\m{log(Mg/H)}=-4.35$. 
The abundance ratio is derived from the metallicity of \ob, $\m{12+log[O/H]}=8.97$, estimated in Section \ref{im}.
We adopt an Mg depletion onto dust of  $-1.3$ dex, that is taken from \citet{2009ApJ...700.1299J} with an assumption of $n(\m{H})\sim1\ \m{cm^{-3}}$.
 The column density of hydrogen is estimated to be $N(\m{H})>2.2\times10^{20}\ \m{cm^{-2}}$.
  It should be noted that Equation (\ref{eq_CD}) can be applied {in the case of one absorbing cloud, but that this equation has also be shown to work adequately for the optical depth $\tau_0<5$.}
  Because the calculated optical depth is $\tau_0>5.9$, this column density estimate would include some systemic uncertainties.
  
We calculate a mass outflow rate, $\dot{M}$, assuming a thin shell geometry. 
\citet{2009ApJ...692..187W} give the mass outflow rate,
 \begin{equation}
 \dot{M}\simeq22\ M_{\odot}\ \m{yr}^{-1}C_f\frac{N(\m{H})}{10^{20}\ \m{cm^{-2}}}\frac{R}{5\ \m{kpc}}\frac{v}{300\ \kms},
 \end{equation}
 where $R$ and $v$ are a radius of the shell wind and the outflow velocity, respectively. 
 We adopt $R\sim7.8\ \m{kpc}$, which is the half of the spatial extent of the [{\sc Oii}] emission in the $NB816_{\m{corr}}$ image.
  If we take $v=220\pm30\ \kms$ that is the average of the outflow velocities derived from Mg{\sc ii} and Fe{\sc ii} lines, the mass outflow rate is estimated to be $\dot{M}>55\pm6\ M_{\odot}\ \m{yr}^{-1}$. 
  The error of the mass outflow rate only includes the uncertainty of the average outflow velocity.
  
  A mass loading factor, $\eta$, characterizes a relation between a mass outflow rate and a SFR, and is defined as $\eta=\dot{M}/\m{SFR}$.
  This mass loading factor serves as a critical function in models of galaxy evolution.
  \citet{2012MNRAS.421.3522H} predict mass loading factors of $0.5-2$ in SF-driven winds by the theoretical study.
Observationally, the mass loading factors of AGNs and non-AGN U/LIRGs are estimated to be $\eta=0.1-1.1$ and $\eta=0.02-0.6$, respectively \citep[][corrected for the Salpeter IMF]{2014arXiv1404.1082A}.
\citet{2005ApJS..160..115R} report the mass loading factors of starburst-dominated galaxies to be $\eta=0.01-1$.
  Using the mass outflow rate lower limit and the {$\m{SFR_{SED}}$\footnote{We use $\m{SFR_{SED}}$, which is the best estimate of \ob's SFR (see Section \ref{mw}). The mass loading factor does not strongly depend on the choice of the SFR.}}, we estimate the mass loading factor of \ob\ to be $\eta>0.8\pm0.1$.
  This mass loading factor is relatively high compared with those of the galaxies of \citet{2014arXiv1404.1082A} and \citet{2005ApJS..160..115R}.

\subsection{Difference of the Velocities of Mg{\sc ii} and Fe{\sc ii}}\label{diff}

The outflow velocities derived from Mg{\sc ii} and Fe{\sc ii} absorption lines are $260\pm40$ and $80\pm50\ \m{\kms}$, respectively. 
Although this difference may not be true due to the marginal $2.7\sigma$ detection of the Fe{\sc ii} line (Section \ref{ss_ldss3}), the difference of these two velocities could be explained by three possibilities. 
First possibility is the emission filling.
As the Mg{\sc ii} transitions are resonantly trapped, these absorption lines are affected by filling from resonance emission lines.
This resonance emission is generated by the foreground gas of the outflow.
 The resonance emission fills the absorption near the systemic redshift.
This emission filling shifts the centroid of Mg{\sc ii} absorption to a bluer wavelength \citep[e.g.,][]{2009ApJ...692..187W,2012ApJ...758..135K,2012ApJ...760..127M,2013ApJ...770...41M}.
However, we cannot conclude whether the emission filling occurs or not, because of the limited signal-to-noise ratio.

Second possibility is the difference of their oscillator strengths, as discussed in \citet{2012ApJ...758..135K}. 
Ionization potentials of Mg ($15.0\ \m{eV}$) and Fe ($16.1\ \m{eV}$) are nearly the same, while the oscillator strengths are different. 
The oscillator strength of the Mg{\sc ii} line at $2796\ \m{\AA}$ ($f_{12}=0.60$) is larger than that of the Fe{\sc ii} line at $2587\ \m{\AA}$ ($f_{12}=0.07$). 
This difference indicates that Mg{\sc ii} is optically thick at a low density where Fe{\sc ii} is optically thin. 
Because of the large cross-section for absorption, Mg{\sc ii} is a tracer of the low-density gas better than Fe{\sc ii}. 
Note that such low-density gas is found far from galaxies, and that the speed of the galactic wind increases with increasing galactocentric radius \citep[e.g][]{2009ApJ...703.1394M,2010ApJ...717..289S,2012MNRAS.426..140D}. 
This physical picture explains that the velocity offset of Mg{\sc ii} absorption line is larger than that of Fe{\sc ii} line. 

Third possibility is the Fe{\sc ii} absorption made by a foreground galaxy.
In Section \ref{res_mos}, we detect the two-component emission lines in our MOSFIRE spectrum. 
If \ob\ is a galaxy merger, the two components would correspond to two merging galaxies.
The velocity offset of the blue component is $-70\pm50\ \kms$, comparable to that of the Fe{\sc ii} absorption line.
Hence, the Fe{\sc ii} absorption could be made in the interstellar medium (ISM) of the foreground galaxy responsible for the blue component.
On the other hand, the velocity offset of Mg{\sc ii} absorption lines is $\Delta v=-260\pm40\ \kms$, which is inconsistent with that of the blue component.
This inconsistency implies that there is no strong Mg{\sc ii} absorption in the foreground galaxy, and that the Mg{\sc ii} absorption lines are made by the outflow of the major \ob\ galaxy.
If it is true, an Mg{\sc ii} absorption of the foreground galaxy is absent, which is puzzling.
However, it is possible that Mg{\sc ii} photons are scattered back into the line-of-sight
from surrounding emission regions with no Mg{\sc ii} absorption.

\subsection{Can the Outflowing Gas Escape from \ob?}\label{escape}
 
Comparing the outflow velocity with the local escape velocity, we examine whether the outflowing gas can escape from the gravitational potential of \ob.
Following the study of \citet{2009ApJ...692..187W}, we estimate the escape velocity, $v_{esc}$, at $r$ \citep[see also][]{2010ApJ...719.1503R,2012ApJ...760..127M}. 
Under the assumption of a singular isothermal halo truncated at $r_h$, the escape velocity is 
\begin{equation}
v_{esc}(r)=v_c\sqrt{2\left[1+\m{ln}\left(\frac{r_h}{r}\right)\right]},
\label{eq_esc}
\end{equation}
where $v_c$ is the circular velocity \citep{1987gady.book.....B}. Although we do not have the $r_h/r$ estimate, this escape velocity very weakly depends on $r_h/r$. 
Assuming the plausible parameter range, $r_h/r=10-100$, we obtain $v_{esc}\simeq(2.6-3.3)v_c$. Thus the escape velocity is $v_{esc}\simeq(3.0\pm0.4)v_c$. The relation between the circular velocity and the velocity dispersion, $\sigma$, of the {\sc [Oii]} emission line is $\sigma\simeq(0.6\pm0.1)v_c$ \citep{1997MNRAS.285..779R,2000AJ....119.1608K,2006ApJ...653.1027W}. Combining this relation, we obtain the escape velocity of $v_{esc}\simeq(5\pm1)\times\sigma$.
 Adopting the velocity dispersion of the {\sc [Oii]} emission lines of \ob, $\sigma=51\pm26\ \kms$, we calculate the escape velocity to be $v_{esc}=250\pm140\ \kms$.
This escape velocity is comparable to the outflow velocity, $v=80-260\ \kms$, implying that some fraction of the outflowing gas can escape from \ob. 
Recently, the analytic models of \citet{2014arXiv1405.3978I} predict the slowly accelerated outflows with increasing radius in the gravitational potential of a cold dark matter halo and a central super-massive black hole.
\citet{2014arXiv1405.3978I} apply their model to the Sombrero galaxy, and the model agrees with the radial density profile measured by observations.
If the outflow is accelerated in \ob\ as claimed by \citet{2014arXiv1405.3978I}, more fraction of the gas would escape from \ob.
The escape of the outflowing gas indicates that the star formation activity may be suppressed in \ob, and that the IGM would be chemically enriched by this outflow process\footnote{Y13 roughly calculate the escape velocity to be $v_{esc}\simeq\sqrt{2GM_{halo}/{r}}$,
 where $M_{halo}$ is the halo mass and $r$ is the radius of the galaxy.  
The escape velocity of \ob\ derived from this equation is $v_{esc}\sim510\ \kms$, if we substitute $M_{halo}=6\times10^{11}\ M_{\odot}$ estimated from the relationship between the stellar and halo masses at $0.74<z<1.0$ of \citet{2012ApJ...744..159L}, and $r=19.2\ \m{kpc}$, which is the Petrosian radius measured in the $Rz$ continuum image with the Petrosian factor of $0.2$.}.

\subsection{Does \ob\ Have an AGN?}\label{agn}

The next question is the energy source of the outflow of \ob\ that would be powered by an AGN activity and/or star formation.
To examine the presence of the AGN, 
we search for a counterpart of {\ob} in the X-ray source catalog provided by \citet{2008ApJS..179..124U}.  
We find no detection within the circle of the X-ray positional error
at the \ob\ position. 
It indicates that \ob\ does not harbor an X-ray luminous AGN with a $2-10\ \m{keV}$ luminosity brighter than $10^{43}$ erg s$^{-1}$. 
However, we cannot rule out the possibility that \ob\ hosts a heavily obscured AGN with a faint X-ray luminosity.



\begin{figure*}
 \begin{minipage}{0.49\hsize}
 \begin{center}
  \includegraphics[width=82mm,bb=20 30 412 412,clip]{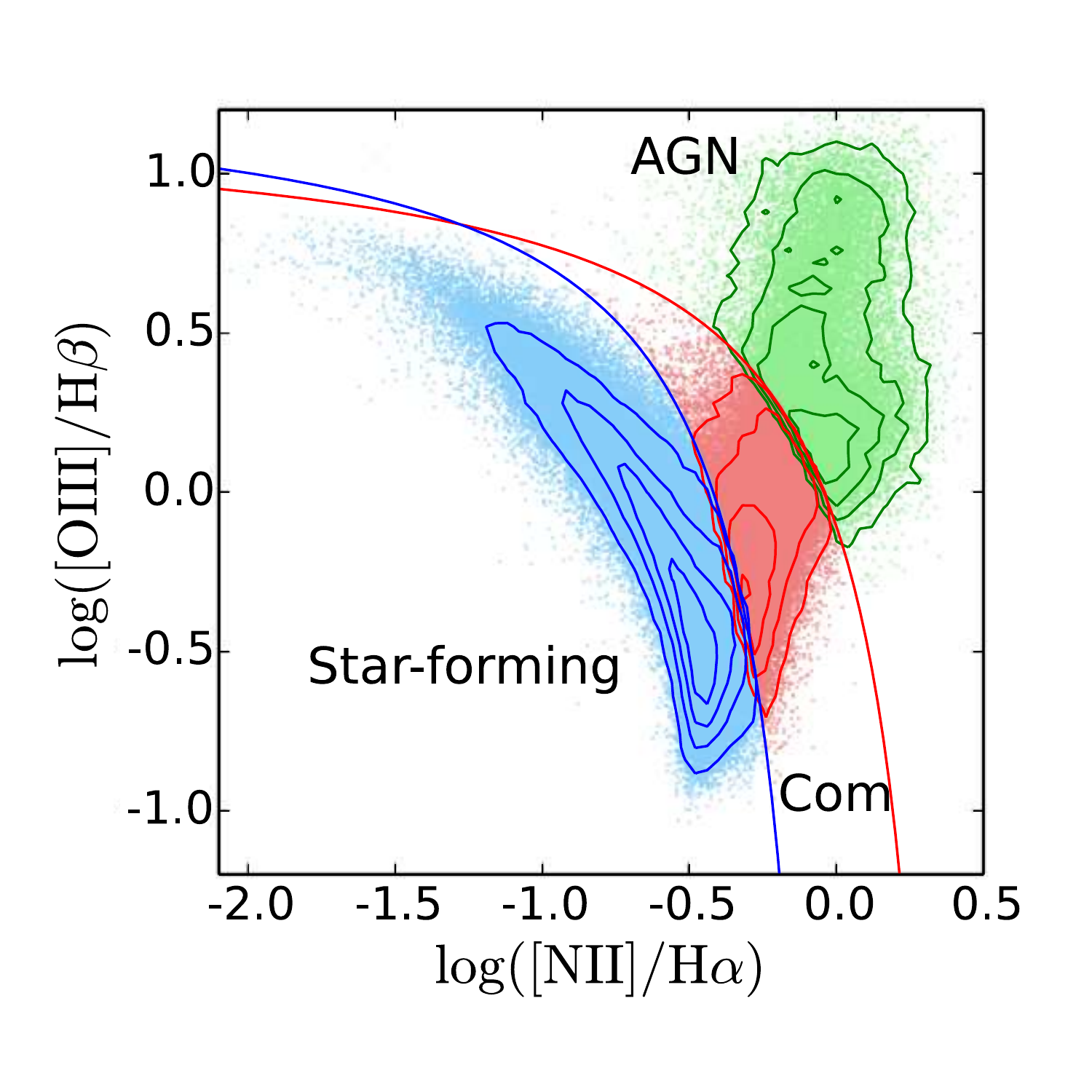}
 \end{center}
  \label{fig:cex}
 \end{minipage}
 \begin{minipage}{0.49\hsize}
 \begin{center}
  \includegraphics[width=82mm,bb=20 30 412 412,clip]{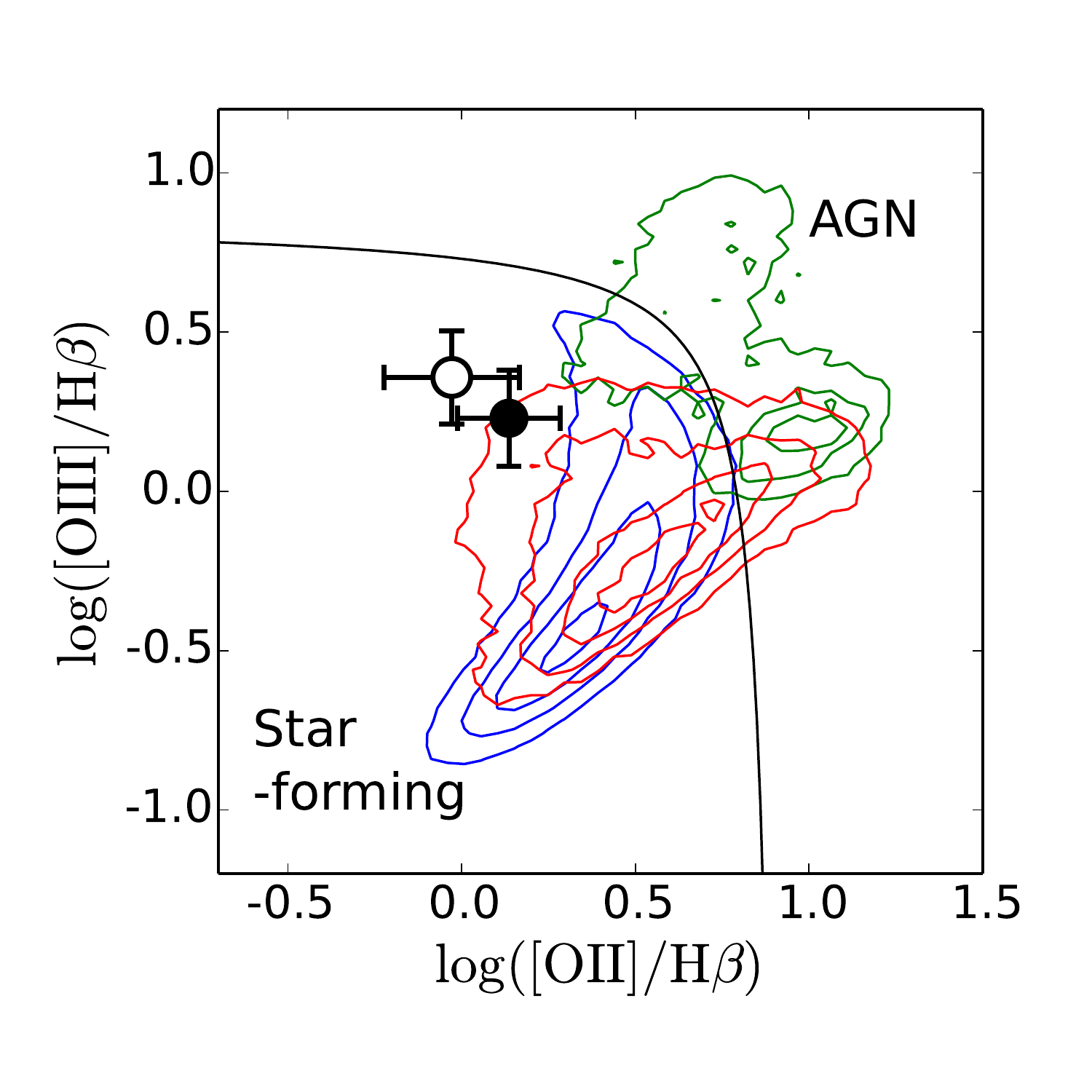}
 \end{center}
  \label{fig:mex}
 \end{minipage}
 
  \begin{minipage}{0.49\hsize}
 \begin{center}
  \includegraphics[width=82mm,bb=20 30 412 412,clip]{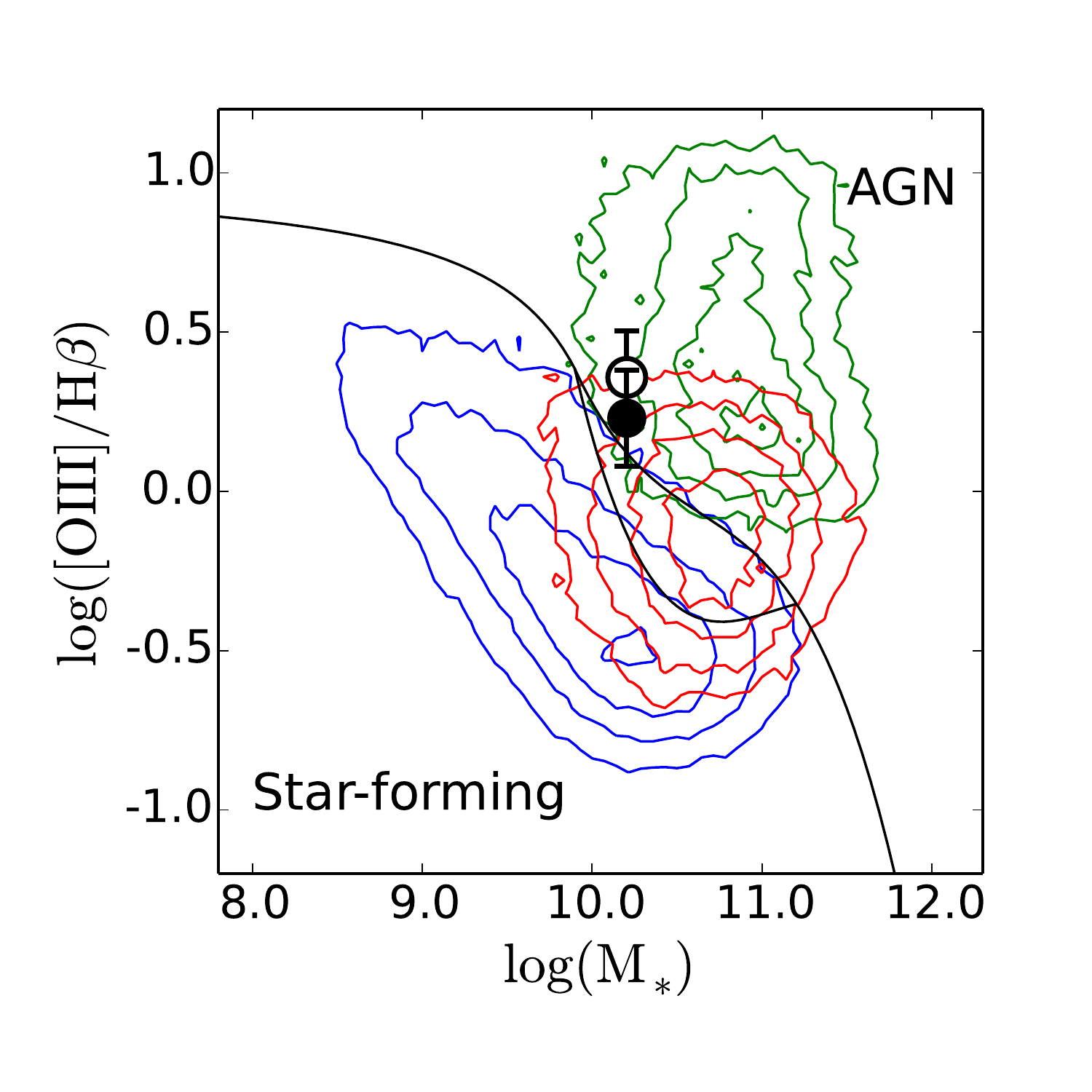}
 \end{center}
  \label{fig:cex}
 \end{minipage}
 \begin{minipage}{0.49\hsize}
 \begin{center}
  \includegraphics[width=82mm,bb=20 30 412 412,clip]{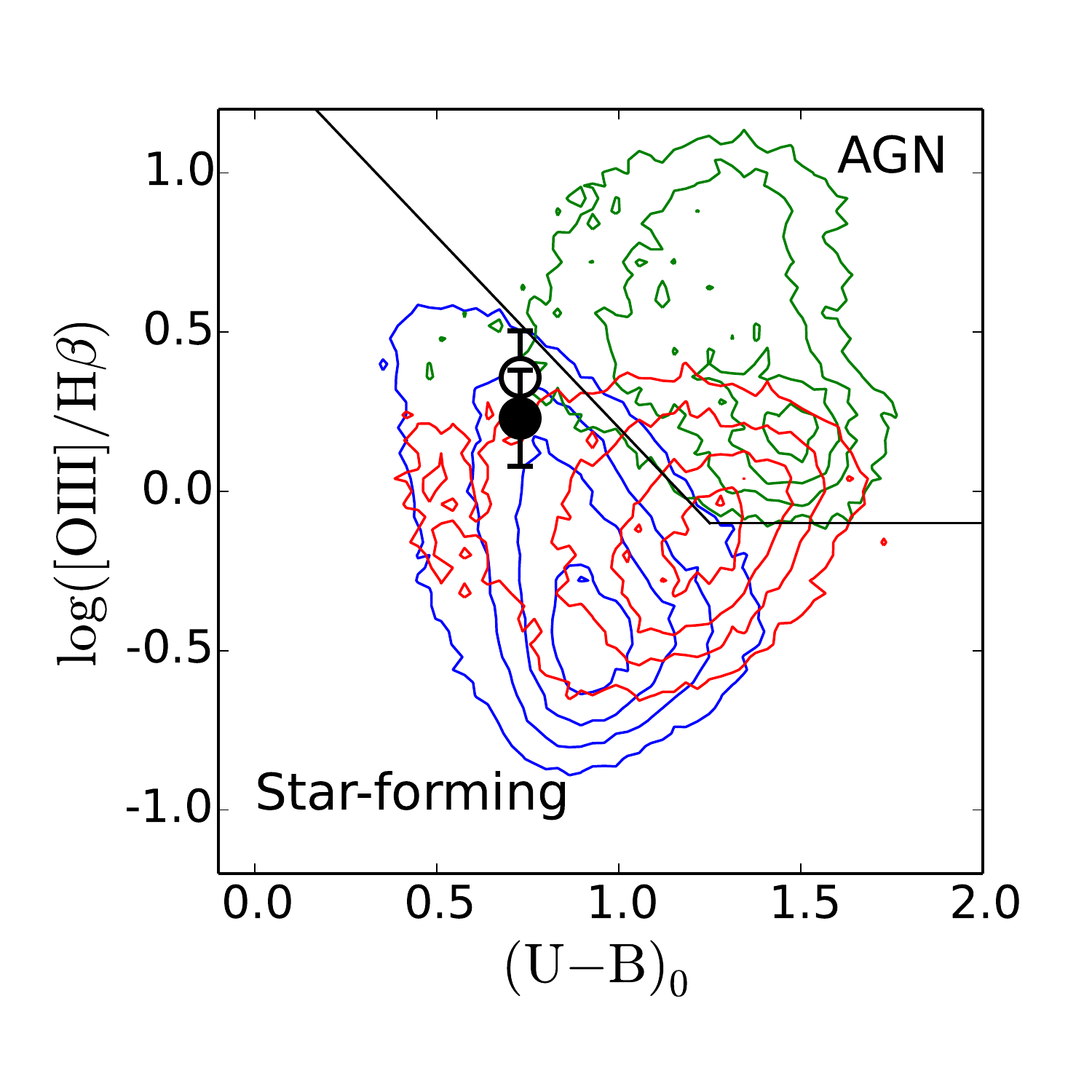}
 \end{center}
  \label{fig:mex}
 \end{minipage}
 
 \caption{BPT, blue, MEx, and CEx diagrams. The upper left panel shows the BPT diagram. {We cannot plot \ob\ in the BPT diagram because the [{\sc Nii}]$\lambda6584$/H$\alpha$ ratio is not available in our study. Yet we use the BPT diagram to classify SDSS galaxies into pure star-forming galaxies, composites, and AGNs, in order to compare them with \ob\ in the blue, MEx, and CEx diagrams.} 
 The blue and red lines are boundaries of pure star-forming galaxies, composites, and AGNs, which are obtained by \citet{2003MNRAS.346.1055K} and \cite{2001ApJ...556..121K}, respectively.
 The dots denote the SDSS galaxies.
The pure star-forming galaxies, composites, and AGNs are shown with the dots and density contours of blue, red, and green colors, respectively.
  The upper right panel is the blue  diagram.
  The black line is a boundary between pure star-forming galaxies and AGNs that is empirically determined by \citep{2010A&A...509A..53L}.
   The lower left panel is the MEx diagram.
   The black lines are boundaries defined by \citet{2011ApJ...736..104J}. 
   The lower right panel shows the CEx diagram.
   The black line is a boundary of \cite{2011ApJ...728...38Y}.
    In the blue, MEx, and CEx diagrams, \ob\ is shown with {the black filled and open circles, which correspond to the extinction-corrected and uncorrected line ratios, respectively. } \label{fig_dia}}
\end{figure*}

The BPT diagram \citep{1981PASP...93....5B} is the diagram of [{\sc Nii}]$\lambda6584$/H$\alpha$ vs. [{\sc Oiii}]$\lambda5007$/H$\beta$, which is widely used to distinguish AGNs and star-forming galaxies. 
 The upper left panel of Figure \ref{fig_dia} shows the BPT diagram. We plot SDSS galaxies taken from the data of SDSS DR7 \citep{2009ApJS..182..543A}\footnote{The emission-line data are taken from the following website: http://www.mpa-garching.mpg.de/SDSS/DR7/}. 
The blue and red classification lines are obtained from \citet{2003MNRAS.346.1055K} and \cite{2001ApJ...556..121K}, respectively.
We classify the SDSS galaxies based on this diagram.
Galaxies that lie below the blue line are classified as pure star-forming galaxies.
Galaxies located between the two classification lines are regarded as composites of an AGN and star-forming regions.
Galaxies above the red line are classified as AGNs.
The pure star-forming galaxies, composites, and AGNs are shown with the dots and contours of blue, red, and green colors, respectively.
Because the [{\sc Nii}]$\lambda6584$ and H$\alpha$ lines of \ob\ are not covered by our spectroscopic data, we cannot plot \ob\ on the BPT diagram.

Instead of the BPT diagram, we adopt other three classification diagrams.
The first is the diagram of the {\sc [Oii]}$\lambda3727/\m{H\beta}$ 
and {\sc [Oiii]}$\lambda5007/\m{H\beta}$ ratios \citep[dubbed "blue diagram"; e.g.,][]{1997MNRAS.289..419R,2004MNRAS.350..396L,2009MNRAS.398...75S,2010A&A...509A..53L}, which substitutes {\sc [Oii]}$/\m{H\beta}$ for [{\sc Nii}]/H$\alpha$.
The blue diagram is presented in the upper right panel of Figure \ref{fig_dia}, and the empirical boundary between pure star-forming galaxies and AGNs \citep{2010A&A...509A..53L} is shown with the black line. 
The second is called the mass-excitation diagram \cite[MEx diagram;][]{2011ApJ...736..104J}, which utilizes the stellar mass instead of [{\sc Nii}]/H$\alpha$.
The lower left panel in Figure \ref{fig_dia} is the MEx diagram with the boundaries in the black lines defined by \citet{2011ApJ...736..104J}.
It should be noted that the boundaries are not well calibrated at $z\gtrsim1$.
The third is the color-excitation diagram \cite[CEx diagram;][]{2011ApJ...728...38Y}, which uses the rest-frame $U-B$ color in place of [{\sc Nii}]/H$\alpha$.
The CEx diagram is shown in the lower right panel.
The black line indicates the boundary defined by \cite{2011ApJ...728...38Y}.
In these three diagrams, contours of the pure star-forming galaxies, composites, and AGNs are plotted with the colors same as the BPT diagram.

In the blue, MEx, and CEx diagrams, \ob\ is denoted by the red circles.
In the blue diagram, we find that \ob\ lies in the composite region.
In the MEx diagram, \ob\ is located not only in the composite region, but also on the edge of the AGN or the pure star-forming galaxy regions.
In the CEx diagram, \ob\ lies on the edge of the AGN or the composite regions, as well as in the pure star-forming region. 
{While it is likely that \ob\ is a composite of an AGN and star-forming regions from these three diagrams, the possibilities of an AGN and a pure star-forming galaxy still remain.}

As described in Section \ref{a}, the FWHM emission line widths of \ob\ are narrow, $70-130\ \kms$. 
It is widely known that 
type-1 AGNs have spectra with very broad permitted lines 
whose FWHM line widths are $\simeq (1-5) \times10^3\ \kms$, 
and with moderately broad ($\sim 500\ \kms$) forbidden lines such as {\sc [Oiii]}$\lambda\lambda4959,5007$. 
The typical FWHM emission line width of type-2 AGNs 
is $500\ \kms$, 
which is similar to those of the forbidden lines in type-1 AGNs. 
Yet \citet{2006MNRAS.372..961K} report that the FWHM emission line widths of type-2 AGNs and composites are $100-600\ \kms$.
Star-forming galaxies typically show line widths narrower than those of AGNs \citep{1989agna.book.....O}. 
{Although the narrow line widths of {\ob} generally imply that \ob\ would be a star-forming galaxy, these line widths are consistent with those of some type-2 AGNs and composites.}

\subsection{Stellar Population of \ob\label{sb}}
\begin{figure}
\begin{center}
\includegraphics[width=80mm,clip,bb=30 30 412 412]{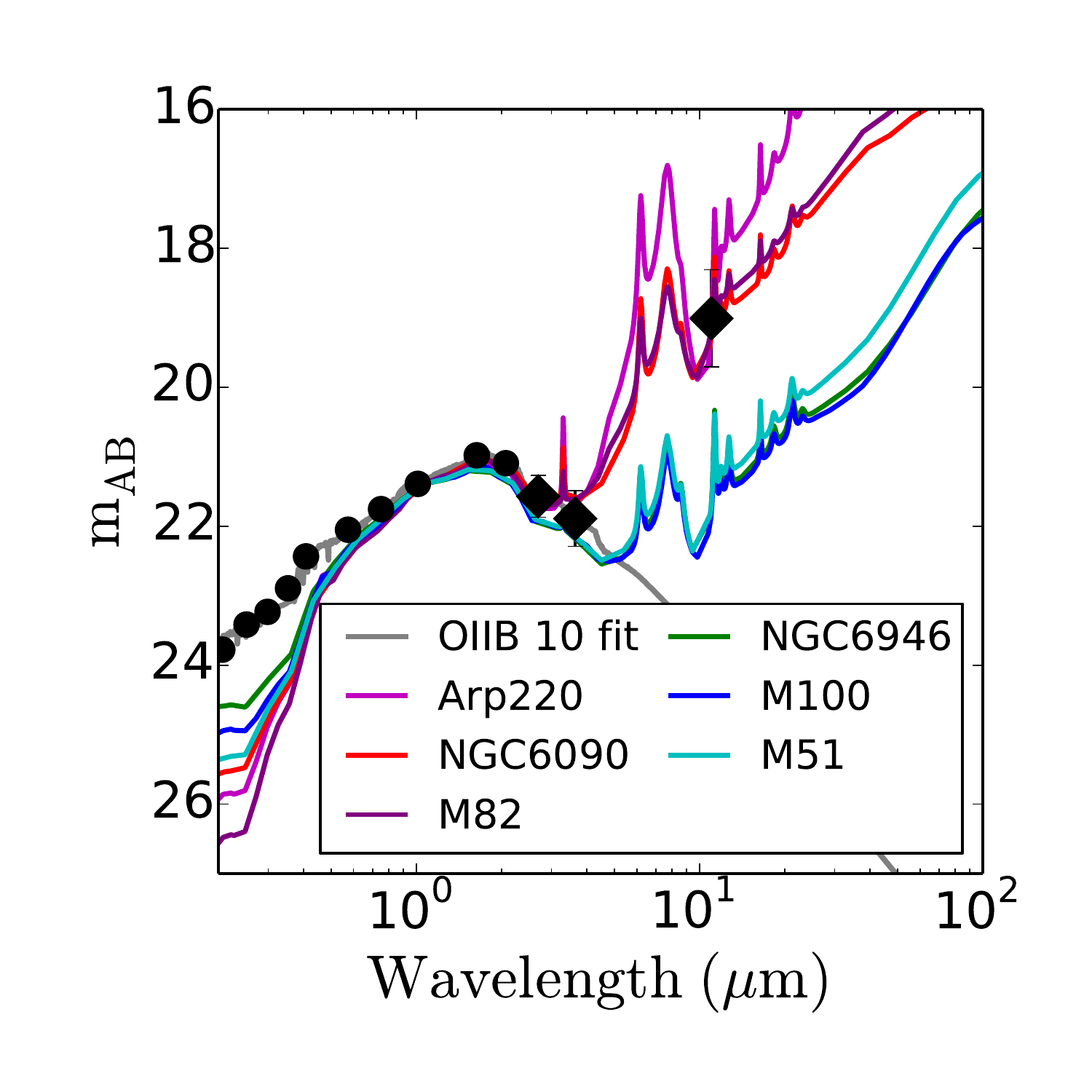}
\end{center}
\caption{SED of \ob\ in the rest frame wavelength. The black diamonds represent the magnitudes obtained in this study, and the black circles show the magnitudes estimated in Y13. The gray line is the result of the SED fitting in Y13. The magenta, red, purple, green, blue, and cyan lines indicate the SEDs of Arp220, NGC6090, M82, NGC6946, M100, and M51 \citep{1998ApJ...509..103S}, respectively. All SEDs are normalized at the $K$ band magnitude. The mid-infrared SED shape of \ob\ is similar to those of the local dusty starbursts, Arp220, NGC6090, and M82, suggesting that the \ob\ is a dusty starburst galaxy. \label{fig_SED}}
\end{figure}

The multi-wavelength SED of \ob\ is shown in Figure \ref{fig_SED}.
Figure \ref{fig_SED} compares the SED of \ob\ with those of local starburst templates \citep{1998ApJ...509..103S}. 
The mid-infrared SED shape of \ob\ is similar to those of \objectname{Arp220}, \objectname{NGC6090}, and \objectname{M82}. 
This indicates that \ob\ is a dusty starburst galaxy, consistent with the non-detection of the Mg{\sc ii} emission line discussed in Section \ref{ao}.

We calculate the SFR of \ob\ using the [{\sc{Oii}}] or H$\beta$ emission line luminosity with the formulae of \citet{1998ARA&A..36..189K}. 
We estimate the $\m{H\alpha}$ luminasity from our extinction-corrected $\m{H\beta}$ luminosity with the line ratio of $L(\m{H\alpha}):L(\m{H\beta})=2.86:1$, under the assumption of Case B recombination in a nebula with a temperature of $T=10000\ \m{K}$ and an electron density of $n_e=100\ \m{cm^{-3}}$ \citep{1989agna.book.....O}. 
The SFRs derived from the {\sc [Oii]} and $\m{H\beta}$ luminosities are ${\m{SFR}_{\textsc{[Oii]}}}=81\pm27$ and ${\m{SFR}_{H\beta}}=96\pm20\ M_{\odot}\ \m{yr}^{-1}$, respectively. 
They are consistent with {the SFR from the SED fitting, ${\m{SFR_{SED}}}=72\pm8\ M_{\odot}\ \m{yr}^{-1}$}. 
We also estimate the SFR with our MIPS $24\ \m{\mu m}$ flux. 
Following the equations in \citet{2009ApJ...692..556R}, we obtain ${\m{SFR_{24\ {{\mu}m}}}}=100\pm70\ M_{\odot}\ \m{yr}^{-1}$.
These SFRs are summarized in Table \ref{tab_sed}.
The SSFR of \ob\  derived from the [{\sc{Oii}}] or H$\beta$ luminosities is $(45-60)\times10^{-10}\ \m{yr}^{-1}$.
This SSFR is higher than those of \objectname{Arp220}, \objectname{NGC6090}, and \objectname{M82}, that are $25$, $1.7$ and $3.1\ \times10^{-10}\ \m{yr}^{-1}$, respectively \citep{1998ApJ...509..103S}.





\subsection{Metallicity and Ionization Parameter\label{im}}
\begin{figure*}
\begin{center}
\includegraphics[width=140mm,clip,bb=30 35 556 412]{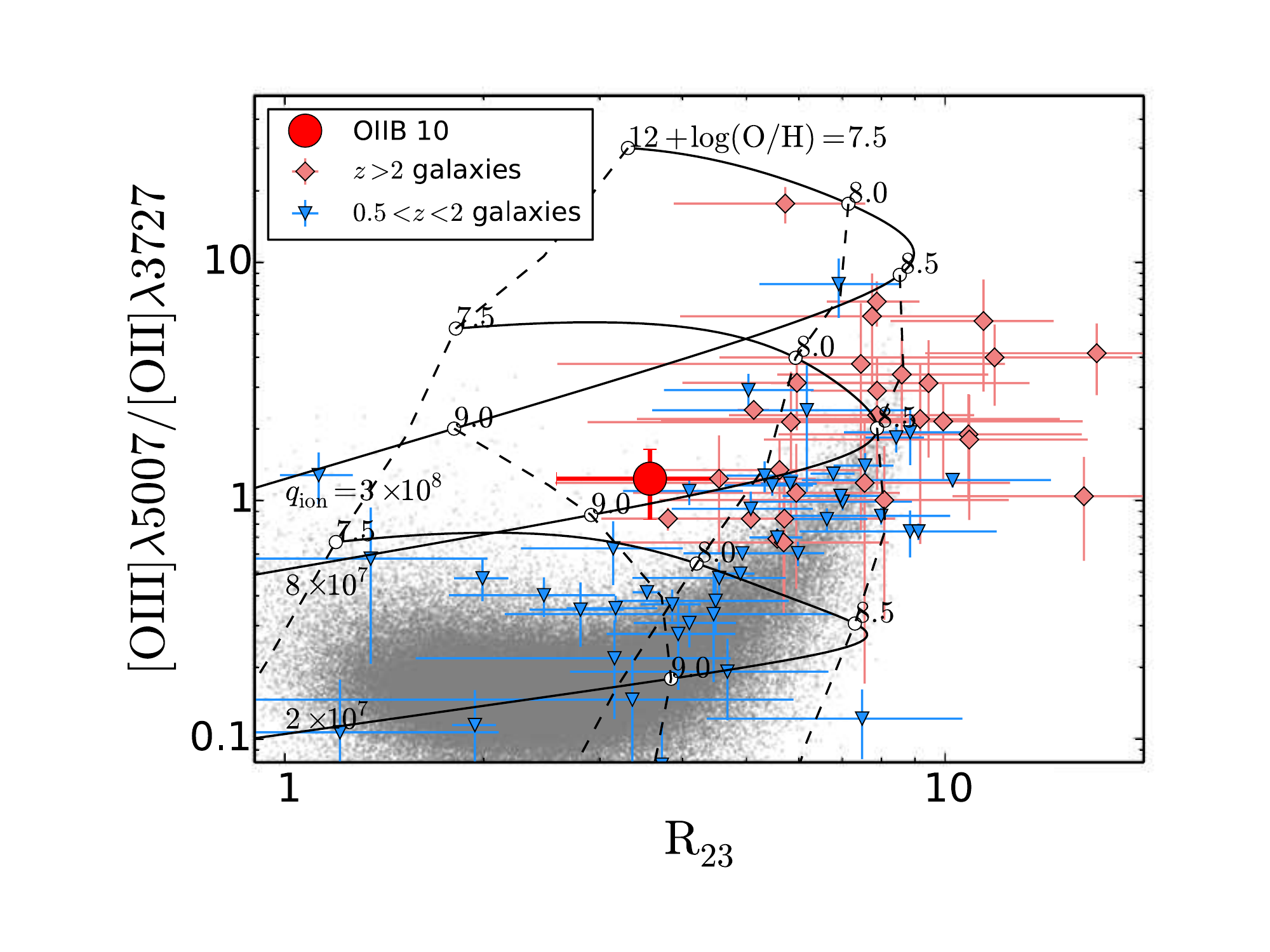}
\end{center}
\caption{$R23$-index vs. {\sc [Oiii]}$\lambda5007$/{\sc [Oii]}$\lambda3727$ diagram. The red circle denotes \ob.
The gray dots represent the SDSS galaxies. The red diamonds and the blue triangles indicate $z>2$ and $0.5<z<2$ galaxies, respectively, that are compiled by \citet{2013arXiv1309.0207N}. 
{The $R_{23}$ and the {\sc [Oiii]}/{\sc [Oii]} values of \ob\ are comparable to those of the $0.5<z<2$ galaxies.}
The black solid curves show the photoionization model tracks with ionization parameters of $q_{\m{ion}}=0.2, 0.8, \m{and}\ 3\times10^8\ \m{cm\ s^{-1}}$, for the metallicity range of $\m{12+log(O/H)}=7.5-9.5$.
The dashed lines indicate the constant metallicities of $\m{12+log(O/H)}=7.5, 8.0, 8.5, \m{and}\ 9.0$.
These model tracks are calculated with the formulae presented in \citet{2002ApJS..142...35K}. 
\label{fig_nakajima}}
\end{figure*}

In this Section, we examine the metal abundance and ionization state of the ISM in \ob.
For simplicity, we assume that \ob\ is dominated by the starburst, and adopt photoionization models.
The $R23$-index and the ratio of {\sc [Oiii]}$\lambda5007$ to {\sc [Oii]}$\lambda3727$ ({\sc [Oiii]}/{\sc [Oii]} ratio) are useful for investigating metallicities, $\m{12+log(O/H)}$, and ionization parameters, $q_{\m{ion}}$, of galaxies \citep{2002ApJS..142...35K,2013ApJ...769....3N,2013arXiv1309.0207N}.
The $R23$-index and {\sc [Oiii]}/{\sc [Oii]} ratio of \ob\ are $R_{23}=3.57\pm0.99$ and $\m{{\textsc{[Oiii]}}/{\textsc{[Oii]}}}=1.24\pm0.40$, respectively.
Figure \ref{fig_nakajima} shows the diagram of $R23$-index vs. {\sc [Oiii]}/{\sc [Oii]} ratio, and \ob\ is denoted by a red circle.
The black solid (dashed) curves indicate photoionization model tracks with the constant ionization parameters (metallicities). 
These model tracks are calculated with the formulae presented in \citet{2002ApJS..142...35K}. 
The SDSS galaxies are plotted with the gray dots.
The red diamonds and the blue triangles represent $z>2$ and $0.5<z<2$ galaxies, respectively, that are compiled by \citet{2013arXiv1309.0207N}.
As shown in Figure \ref{fig_nakajima}, the $R_{23}$ value and the {\sc [Oiii]}/{\sc [Oii]} ratio of \ob\ are comparable to those of the $0.5<z<2$ galaxies.


Following the equations in \citet{2004ApJ...617..240K}, we estimate the metallicity and ionization parameter of \ob\ from the $R23$-index and {\sc [Oiii]}/{\sc [Oii]} ratio.
There are two solutions: $(\m{12+log[O/H]},\ q_{\m{ion}} [10^8\ \m{cm\ s^{-1}}])$ $=$ $(8.97\pm0.08,\ 1.3\pm0.7)$ and $(7.87\pm0.19,\ 0.32\pm0.08)$, which are high and low metallicity solutions, respectively.
The SED fitting results presented in Section \ref{mw} reveal that the stellar mass of \ob\ is relatively high, 
$1.6\times10^{10}\ M_{\odot}$.
We assume that this high stellar mass is due to the past star formation that largely contributes to the metal enrichment of ISM. 
Moreover, as discussed in Section \ref{sb}, \ob\ is a dusty starburst galaxy. 
Thus, \ob\ should have the high-metallicity solution, $(\m{12+log[O/H]},\ q_{\m{ion}} [10^8\ \m{cm\ s^{-1}}])$ $=$ $(8.97\pm0.08,\ 1.3\pm0.7)$, which indicates the high ionization parameter.
The ionization parameter of \ob\ is comparable to those of the $z>2$ galaxies.
However, the metallicity of \ob\ is significantly higher than those of the $z>2$ galaxies, and similar to those of SDSS metal-rich galaxies in Figure \ref{fig_nakajima}. 




\subsection{Comparison with Other Galaxies}
We compare the properties of \ob\ with those of other {\sc Oii}Bs.
Y13 study {\sc Oii}B\ 1, 4, and 8 with the optical spectroscopic data, and find that {\sc Oii}B\ 1 and 4 have outflow signatures. 
The outflow velocity of \ob\ is comparable to that of {\sc Oii}B\ 4 ($\sim\ 200\ \m{\kms}$), and {much less than} that of {\sc Oii}B\ 1 ($500-600\ \m{\kms}$) that exhibits the AGN activity. 
Mg{\sc ii}$\lambda\lambda$2796,2804 and Fe{\sc ii}$\lambda$2587 absorption lines, which are found in \ob, are also detected in {\sc Oii}B\ 4.
{In {\sc Oii}B\ 1, however, only the Fe{\sc ii}$\lambda$2587 absorption line is marginally detected.}
These facts imply that the outflow mechanism of \ob\ may be similar to that of {\sc Oii}B\ 4. 
The ionization parameter of \ob\ is $q_{\m{ion}}=(1.3\pm0.7)\times10^8\ \m{cm\ s^{-1}}$, corresponding to $\m{log}U=-2.36\pm0.26$. This ionization parameter is lower than that of {\sc Oii}B\ 1, $\m{log}U>-1.0$, which is estimated from the {\sc [Oii]}$\lambda3727$/Fe{\sc ii}{$^\ast$}$\lambda2613$ ratio.


{Then, we compare \ob\ with typical star-forming galaxies at $z\sim1.5$ and $2.2$.
\citet[hereafter M14]{2014arXiv1402.0510M} study the emission line galaxies at $z\sim1.5$ and $2.2$ with near-infrared spectroscopic data.
Their galaxies have strong [{\sc Oiii}] and/or H$\alpha$ emission lines.
Thus, the average spectrum of these galaxies in M14 represents the typical star-forming galaxies.
The [{\sc Oiii}] line width of \ob\ presented in Table \ref{tab_el} is comparable to that of the average spectrum in M14, $75\ \m{\kms}$.
The ionization parameter of \ob, $\m{log}U=-2.36\pm0.26$, is also simililar to that of the average spectrum in M14, $\m{log}U=-2.49^{+0.07}_{-0.14}$.
However, the metallicity of \ob, $\m{12+log[O/H]}=8.97\pm0.08$, is higher than that of the average spectrum in M14, $\m{12+log[O/H]}=8.37$.
This indicates that \ob\ is more metal-rich than the typical star-forming galaxies at $z\sim1.5$ and $2.2$.}

\subsection{What is \ob?}\label{what}

We discuss the physical origin of \ob.
As described in Section \ref{res_mos}, we identify the two-components emission lines in \ob{, and the red components appear to be spatially more extended than the blue components.}
The origin of the two components is unknown, but there are three possibilities: (1) two strong star-forming regions, (2) a galaxy merger, and (3) a combination of a galaxy and an outflow knot. 
The first possibility is that \ob\ has two strong star-forming regions, which make the two emission line components. 
The velocity difference of the two components ($170\pm50\ \kms$) may be due to a rotation or an infall of these two star-forming regions.
The second possibility is that the \ob\ is a merger system, and that the two components correspond to the two galaxies.
When galaxies merge, large amounts of the ISM fall on the central regions. 
This leads to the starburst processes. 
As discussed in Section \ref{sb}, \ob\ is a dusty starburst galaxy, which supports this merger-triggered starburst scenario.
Generally, merging galaxies have broad ($v_{\m{FWHM}}\agt 350\ \kms$) and narrow ($v_{\m{FWHM}}\alt 350\ \kms$) emission lines that are produced by a shocked gas and {\sc Hii} regions, respectively \citep[e.g.,][]{2011ApJ...734...87R,2012ApJ...757...86S,2013ApJ...768...75R}. 
However, we do not detect broad lines in \ob.
The non-detection of the broad lines implies that the shock excitation would not be dominant in \ob.
The third possibility is that {\sc Hii} regions and one collimated outflow knot of the ionized gas are responsible for the two components, respectively.
As discussed in Section \ref{ao}, \ob\ has the outflow.
If the outflow is beamed to us, the faint blue component of the emission line would be the outflow knot.
In this case, the bright red component is originated from the {\sc Hii} regions.
 The outflow velocity should be comparable to the velocity difference of the two emission line components, $170\pm50\ \m{\kms}$ (Section \ref{res_mos}).
With $z_{\m{R}}$ and the absorption line velocities, we estimate the outflow velocity for the red component to be $160-390\ \kms$,
comparable to the velocity difference of the two components.
{None of these three possibilities can be conclusively ruled out given current observational results including the fact that the red components appear to be more extended than the blue components.}

The next question of \ob\ is the energy source of the outflow found in Section \ref{ao}.
From the diagrams of Figure \ref{fig_dia}, it is likely that \ob\ is a composite of an AGN and star-forming regions.
The narrow emission line widths of \ob, $70-130\ \kms$ (Section \ref{a}), are comparable not only to star-forming galaxies but also to type-2 AGNs and composites.
Thus a type-2 AGN and star formation would drive the outflow of \ob.
{For further investigation of the outflow and the presence of an AGN, observations with adaptive optics (AO) would be needed.}

\section{Summary}\label{s}
We present the Keck/MOSFIRE and Magellan/LDSS3 spectroscopy and the archival imaging data of thirteen bands for the object with the spatially extended {\sc [Oii]}$\lambda\lambda3726,3729$ emission at $z=1.18$, {\sc [Oii]} blob 10 (\ob). Following the study of \citet{2013ApJ...779...53Y}, our data provide new insight into the physical properties of \ob. 
This is the first detailed spectroscopic study {of} oxygen-line blobs {which includes the analyses of the escape velocity, the mass loading factor, and the presence of an AGN}, and is a significant step
for understanding the nature of oxygen-line blobs and the relationship between gas outflow and star formation
quenching at high redshift.
The major results of our study are summarized below.

\begin{enumerate}

\item By our MOSFIRE observations, we identify H$\beta$ and {\sc [Oiii]}$\lambda\lambda4959,5007$ emission lines, all of which show the profiles of the two components that are called blue and red components. 
The average redshift of all components is $z_{\m{sys}}=1.1800\pm0.0002$, which is defined as the systemic redshift of \ob.
The velocity difference of the blue and red components is $170\pm50\ \kms$.
We estimate the average FWHM line widths of the blue and red components to be $\m{FWHM_B}=90\pm50\ \kms$ and $\m{FWHM_R}=120\pm40\ \kms$, respectively.

\item We detect {\sc [Oii]}$\lambda\lambda3726,3729$ emission lines in our LDSS3 VPH-red grism spectrum.
Although we identify the doublet lines of {\sc [Oii]}, the blue and red components are not resolved in our spectrum.
The redshift of {\sc [Oii]} lines is $z=1.1800\pm0.0007$, consistent with the systemic redshift.
The FWHM line width is estimated to be $120\pm60\ \kms$.

\item In our LDSS3 VPH-blue grism spectrum, we identify blueshifted Mg{\sc ii}$\lambda\lambda$2796,2804 and Fe{\sc ii}$\lambda$2587 absorption lines.
The velocity offsets from the systemic velocity are $\Delta{v}=-260\pm40\ \m{\kms}$ and $\Delta{v}=-80\pm50\ \m{km\ s^{-1}}$ for  the Mg{\sc ii} and Fe{\sc ii} absorption lines, respectively.
These blueshifted absorption lines indicate that \ob\ has an outflow whose velocity is $80-260\ \kms$.

\item The escape velocity of \ob\ is estimated to be $v_{esc}=250\pm140\ \kms$.
 The outflow velocity of \ob, $80-260\ \kms$, is comparable to this escape velocity, implying that some fraction of the outflowing gas would escape from \ob.
This indicates that the star formation activity could be suppressed by the outflow, and that the chemical enrichment of IGM may take place. 

\item To examine the presence of the AGN, we investigate line ratios, the stellar mass, and the color with the "blue", MEx, and CEx diagrams.
These diagrams indicate that \ob\ {would be} a composite of an AGN and star-forming regions, {but} do not rule out the possibilities of an AGN and a pure star-forming galaxy.
{While the narrow line widths of \ob\ are suggestive of a star-forming galaxy, they are also consistent with an AGN or a composite.}

\item The SED of \ob\ suggests that \ob\ is a dusty starburst galaxy, because the mid-infrared SED shape is similar to those of local starburst galaxies.
We estimate the metallicity and ionization parameter of \ob\ to be $\m{12+log(O/H)}=8.97\pm0.08$ and $q_{\m{ion}}=(1.3\pm0.7)\times10^8\ \m{cm\ s^{-1}}$, respectively.

\item The two-component emission lines found in our MOSFIRE spectrum would indicate three possibilities: (1) two strong star-forming regions, (2) a galaxy merger, and (3) a combination of a galaxy and an outflow knot.
These three possibilities are all consistent with the results from our observations.
The outflow may be driven by a composite of star formation and a type-2 AGN.

\end{enumerate}



\acknowledgments

We thank the anonymous referee for careful reading and valuable comments that improved clarity of the paper.
We are grateful to Malte Schramm for his efforts on taking K-band imaging data of OIIB 10
with Subaru/IRCS-AO188.
We thank Akio Inoue, Masao Mori, Kentaro Nagamine, Tomoki Saito, Kazuhiro Shimasaku, and Masayuki Umemura for useful comments and discussions.
We appreciate the staff of the Keck Observatory for their help with the observations.
We wish to recognize and acknowledge the very significant cultural role
and reverence that the summit of Mauna Kea has always
had within the indigenous Hawaiian community. We are
most fortunate to have the opportunity to conduct observations from this mountain.
We thank the exquisite support of the staff at Las Campanas.
 This work was supported
by KAKENHI (23244025)
Grant-in-Aid for Scientific Research (A) through Japan Society
for the Promotion of Science (JSPS).
S.Y. and K.N. acknowledge
the JSPS Research Fellowship for Young Scientists.

\end{document}